\documentclass[prd,showpacs,preprintnumbers,amsmath,amssymb]{revtex4}

\usepackage{graphicx}
\usepackage{dcolumn}
\usepackage{bm}
\def\simge{
    \mathrel{\rlap{\raise 0.511ex
        \hbox{$>$}}{\lower 0.511ex \hbox{$\sim$}}}}
\def\simle{
    \mathrel{\rlap{\raise 0.511ex
        \hbox{$<$}}{\lower 0.511ex \hbox{$\sim$}}}}

\def\beqn{\begin{equation}}
\def\eeqn{\end{equation}}
\def\barr{\begin{eqnarray}}
\def\earr{\end{eqnarray}}
\def\bc{\begin{center}}
\def\ec{\end{center}}

\begin{document}

\title{
First lattice study of low-energy charmonium-hadron interaction
}

\author{Kazuo~Yokokawa$^{(1)}$, Shoichi~Sasaki$^{(1,2)}$, Tetsuo~Hatsuda$^{(1)}$ and Arata~Hayashigaki$^{(3)}$}
\address{$^{(1)}$ Department of Physics, The University of Tokyo, Tokyo 113-0033, Japan}
\address{$^{(2)}$ RIKEN BNL Research Center, Brookhaven National Laboratory, Uptopn, NY 11973, USA}
\address{$^{(3)}$
Institut f\" ur Theoretische Physik, J.W.~Goethe Universit\" at, D-60438 Frankfurt, Germany}

\date{\today}
\begin{abstract}
We study the scattering lengths
of charmonia ($J/\psi$ and $\eta_c$) with light hadrons ($\pi$, $\rho$ 
and $N$) 
by the  quenched lattice QCD simulations 
on $24^3 \times 48$, $32^3 \times 48$ and
$48^3 \times 48$ lattices with the lattice spacing
$a \simeq 0.068 $ fm. The scattering length is
extracted by using the L\"{u}scher's phase-shift
formula together with the measurement of
the energy shift $\Delta E$  of two hadrons on the lattice.
We find that there exist attractive interactions 
in all channels,
$J/\psi(\eta_c)$-$\pi$,
$J/\psi(\eta_c)$-$\rho$
and $J/\psi(\eta_c)$-$N$:
The $s$-wave $J/\psi$-$\pi$ ($\eta_c$-$\pi$) scattering length is determined
as 0.0119$\pm$0.0039 fm (0.0113$\pm$0.0035 fm) and the corresponding
elastic cross section at the threshold
becomes ${0.018}^{+0.013}_{-0.010}$ mb 
(${0.016}^{+0.011}_{-0.008}$ mb).
Also, the $J/\psi$-$N$ ($\eta_c$-$N$) spin-averaged scattering length  
is  0.71$\pm$0.48 fm (0.70$\pm$0.66 fm), which is 
at least an order of magnitude larger than the charmonium-pion scattering length.
The volume dependence of the energy shifts is also 
investigated to check the expected $1/L^3$ behavior of 
$\Delta E$ at a large spatial size $L$.
 \end{abstract}


\maketitle


\section{Introduction}
\label{sec:introduction} 
 
 Properties of single hadrons
 have been studied in quantum chromodynamics (QCD) by using 
 various techniques such as the QCD sum rules and lattice QCD simulations.
 On the other hand,  the interaction between color-singlet hadrons
 is not fully explored yet because of their complex nature originating from 
 quark exchanges and multiple gluon exchanges.

 In this paper, we study the low-energy elastic scattering of 
 charmonia ($J/\psi$ and $\eta_c$)
 with iso-nonsinglet hadrons composed of up and down quarks 
($\pi$, $\rho$ and $N$) on the basis of the 
 quenched lattice QCD simulations.
 Since the valence quarks in each hadron in the initial state
 stay in  the same hadron in the final state,  the  process we consider is
 much simpler than the interactions between light-hadrons, while
 it is still non-trivial in the sense that non-perturbative
 gluon exchanges play essential roles.
 Furthermore,  such interaction has a direct relation to 
 the physics of charmonium-nucleus ($A$) bound states~\cite{BST,Wasson,LMS,BG97,TEO}
 and an indirect relation to the   
 the elastic and inelastic charmonium-hadron interactions 
 at high energies~\cite{RSZ,STT,STST}.
 
The charmonium-$N$ interaction at low energies has been
discussed in the framework of operator product expansion 
~\cite{BP,Dima,AH,WE,SV,SL} and in some hadronic model~\cite{STT}.
For example, Hayashigaki has shown  that the 
$s$-wave $J/\psi$-$N$ scattering length $a_0^{J/\psi\textit{\rm -}N}$ is about  0.1 fm by
 using QCD sum rules \cite{AH}, 
while Brodsky and Miller found that it is about 0.25 fm from the 
 gluonic van der Waals interaction~\cite{BG97}.
Recently, Sibirtsev and Voloshin suggested that a lower bound of the 
$a_0^{J/\psi\textit{\rm -}N}$ is as large as 0.37 fm from the 
 multipole expansion analysis of chromo-polarizability~\cite{SV}.
In the above-mentioned calculations, the elastic cross section at the threshold 
reads 1.3 mb, 7.9 mb and 17 mb, respectively.
Brodsky, Schmidt and de T\'eramond introduced an
 effective charmonium-nucleon potential
$V(r)= -\alpha {\rm e}^{-\mu r}/r$ with the parameters
$\alpha = 0.42 \sim 0.59$ and $\mu =$ 6.0 GeV estimated by
 a phenomenological model of Pomeron interactions~\cite{BST}.
 They argued that the $J/\psi$-$A$ bound system may be realized for mass number $A \ge 3$
 if the attraction is sufficiently large, which was later confirmed
 by Wasson, who solved  the Schr\"odinger equation for the charmonium-nucleus system~\cite{Wasson}.

 Unlike the case of $J/\psi$-$N$, 
 the interaction of the charmonium with the pion has a special
 feature due to the Nambu-Goldstone nature of the pion.
  The current algebra shows that
 the  $s$-wave scattering length of a heavy hadron ($H$) 
 with the pion is given by the well-known formula~\cite{SW_SL}:
\begin{eqnarray}
a_0^{H\textit{\rm -}\pi} 
= -\left( 1+\frac{M_{\pi}}{M_H} \right)^{-1}
\frac{M_{\pi}}{4\pi f_\pi^2} \vec{I}_\pi \cdot \vec{I}_H +O(M_\pi^2), 
\label{ChPT}
\end{eqnarray}
where $\vec{I}_{\pi}$ ($\vec{I}_H$) is the isospin vector of $\pi$ ($H$) and 
 $ \vec{I}_\pi \cdot \vec{I}_H = \frac{1}{2} \left[I(I+1) -I_H(I_H+1)-2 \right]$
with $I$ being the total isospin of the $\pi$-$H$ system. 
For  the $\pi$-$N$ case, this is known as the Tomozawa-Weinberg  relation. 
 Eq.~(\ref{ChPT}) indicates that the soft pion decouples
 from any hadrons in the chiral limit ($M_{\pi} \rightarrow 0$).
  More importantly in our context,  
  the leading term in Eq.~(\ref{ChPT}) 
  vanishes even if $M_{\pi}$ is finite as long as  $H$
 is  iso-scalar (such as $J/\psi$ and $\eta_c$):
 Namely, the low-energy pion interaction with 
 the charmonium states is very weak of $O(M_{\pi}^2$).
 An alternative estimate based on the color-dipole description 
of  heavy quarkonia shows a small but attractive interaction between 
 the charmonium and the pion \cite{BP,Dima,FK}. Here, the contribution 
 stems from the QCD trace anomaly and the effect starts from $O(M_\pi^2) $ as consistent
with the current algebra analysis~\cite{CS}.

 The method we employ for extracting the charmonium-hadron scattering lengths
 is the quenched lattice QCD simulations together with the 
  phase-shift formula by L\"{u}scher \cite{ML}.
The basic idea is to put two hadrons in a finite box with a spatial size 
$L$ and to measure the energy of interacting hadrons relative 
to the energy of the non-interacting hadrons.
Such an energy shift $\Delta E$ can be translated into the 
scattering  phase shift at low energies and hence
to the $s$-wave scattering length in the limit of zero relative momentum.   
Applications of  L\"uscher's formula to numerical simulations in lattice QCD 
have been previously done for $\pi$-$\pi$, $\pi$-$N$ and $N$-$N$ systems 
in  Refs.~\cite{{SimSL1}, {SimSL2}, {Krms}, {I2PiPi}, {NPLQCD}} and
also for a hypothetical bound state system in Ref.~\cite{SS}.
 Our lattice data show that the low-energy interaction of the 
 charmonia with light hadrons is always attractive.  
 Among others, the charmonium-$\pi$ interaction is
  quite small in accordance with the current algebra result, while
  the  charmonium-$\rho$ and  charmonium-$N$ interactions are likely
  to be large and may even support the 
   charmonium-nucleus bound states according to the phenomenological
    analyses mentioned above.

This paper is organized as follows.
In Sec.~II, we recapitulate the method to calculate 
the $s$-wave scattering phase shift
in lattice QCD  and discuss 
its application of the L\"uscher's phase-shift formula
to the  case of attractive interactions.
In Sec.~III, details of our Monte Carlo simulations and 
numerical results of $J/\psi$-hadron and $\eta_c$-hadron interactions
at low energy are given. Physical implications
 of our results are also discussed.
Finally, Sec.~IV is devoted to summary and concluding remarks. 
In Appendix A, some details of the spin projections for
$J/\psi$-$\rho$ and $J/\psi$-$N$ systems are described.

\section{Two Hadrons in a finite box}
 \label{def-IntOp}
It has been shown by L\" uscher that the $s$-wave scattering phase shift 
is related to the energy shift $\Delta E$ in the total energy
of two hadrons in a finite box ~\cite{ML}. 
To measure the total energy of two hadrons ($h_1$ and $h_2$) 
in the center-of-mass frame, we define the following four-point correlation function:
\begin{equation}
G^{h_1\textit{\rm -} h_2}(t_4, t_3; t_2, t_1)=
\left\langle 
{\cal O}^{h_2}(t_4){\cal O}^{h_1}(t_3)
\left(
{\cal O}^{h_2}(t_2){\cal O}^{h_1}(t_1)
\right)^{\dagger}
\right\rangle,
\end{equation}
where each hadron is projected onto  the zero momentum state
by the summation over all spatial coordinates $\vec x$, i.e. 
${\cal O}^{h}(t)=\sum_{\vec x} {\cal O}^{h}({\vec x}, t)$.
To avoid the Fierz re-arrangement of two-hadron
operators, we choose $t_4=t_3+1$ and $t_2=t_1+1$.
Let us first consider here the scattering processes, 
$J/\psi$-$\pi$, $J/\psi$-$\rho$ and $J/\psi$-$N$
(Cases for $\eta_c$-$\pi$, $\eta_c$-$\rho$ and $\eta_c$-$N$
  will be discussed in Sec.~\ref{sec:etac}).
Accordingly, the four-point correlation functions are defined by
\begin{eqnarray}
G^{J/\psi \textit{{\rm -}} \pi}_{ij}(t, t_{\rm src})    &=&  
\left\langle 
{\cal O}^{\pi}(t+1)  {\cal O}^{J/\psi}_i(t)
\left({\cal O}^{\pi}(t_{\rm src}+1)  {\cal O}^{J/\psi}_j(t_{\rm src}) \right)^\dagger
\right\rangle,  \\
G^{J/\psi \textit{{\rm -}} \rho}_{ij; kl}(t,t_{\rm src}) &=&
\left\langle 
{\cal O}^{\rho}_i(t+1)  {\cal O}^{J/\psi}_j(t)
\left({\cal O}^{\rho}_k(t_{\rm src}+1)  {\cal O}^{J/\psi}_l(t_{\rm src}) \right)^\dagger
\right\rangle,    
\label{Eq:FourPointRho}
\\
G^{J/\psi \textit{{\rm -}} N}_{ij}(t,t_{\rm src})     &=&   
\left\langle 
{\cal O}^{N}(t+1)  {\cal O}^{J/\psi}_i(t)
\left({\cal O}^{N}(t_{\rm src}+1)  {\cal O}^{J/\psi}_j(t_{\rm src}) \right)^\dagger
\right\rangle .
\label{Eq:FourPointNuc}
\end{eqnarray}
Here we use the conventional interpolating operators,
\begin{eqnarray}
{\cal O}^{J/\psi}_\mu({\vec x}, t) &=& \bar{c}_a({\vec x}, t) \gamma_\mu c_a({\vec x}, t),\\
{\cal O}^\pi({\vec x}, t)  &=& \bar{u}_a({\vec x}, t) \gamma_5 d_a({\vec x}, t),\\
{\cal O}^\rho_\mu({\vec x}, t) &=&\bar{u}_a({\vec x}, t) \gamma_\mu d_a({\vec x}, t),\\
{\cal O}^{N}({\vec x}, t) &=& \epsilon_{abc} \Big[ u^T_a({\vec x}, t) C\gamma_5 d_b({\vec x}, t)\Big]
u_c({\vec x}, t),
\end{eqnarray}
for $J/\psi$, $\pi$, $\rho$ and $N$, respectively.
$C$ is the charge conjugation matrix, $C=\gamma_4\gamma_2$. 
Also, $a,b$ and $c$ are color indices, $i, j, k$ and $l$ are spatial Lorentz indices, 
and $u$, $d$ and $c$ are the up, down and charm quark fields.
For the $s$-wave $J/\psi$-$\pi$ scattering, the total spin is restricted to be 1.
 Then we simply choose diagonal correlation averaged over the spatial indices,
 $\frac{1}{3}\sum_{i}G^{J/\psi \textit{\rm -} \pi}_{ii}(t, t_{\rm src})$,
  to  extract the scattering length.
On the other hand, for $s$-wave $J/\psi$-$\rho$ and $J/\psi$-$N$ scatterings,  
different total spin states are allowed: Spin-0, 1 and 2 states for $J/\psi$-$\rho$ and
spin-1/2 and 3/2 states for $J/\psi$-$N$. 
Therefore, we need  appropriate spin projections to disentangle each
spin contribution from four-point correlation functions.
Details of such spin projection are described in Appendix A.

The hadronic two-point functions are defined as 
$G_{h}(t, t_{\rm src}) =\left\langle {\cal O}^{h}(t)   
{\cal O}^{h\dagger}(t_{\rm src})\right\rangle$. 
For the vector mesons such as $J/\psi$ and $\rho$,
we take an average over the spatial Lorentz indices as 
$\frac{1}{3}\sum_{i=1,2,3}\left\langle {\cal O}_{i}^{h}(t)   
{\cal O}_{i}^{h\dagger}(t_{\rm src})\right\rangle$ 
so as to obtain possible reduction of statistical errors.
Note that we have neglected  the disconnected diagrams 
(such as self-annihilation of $J/\psi$) in  evaluating the two-point function of $J/\psi$
 and the four-point functions with $J/\psi$ in our simulations.
 Contributions from the disconnected diagrams in the 
vector channel are known to be negligibly small for strange and 
charm quarks in numerical simulations \cite{UK,TARO}
 in accordance with the mechanism of Okubo-Zweig-Iizuka (OZI) suppression.

Let us now define a ratio $R^{J/\psi\textit{\rm -}h}(t)$ to extract the 
total energy of two hadrons ($E_{J/\psi\textit{\rm -}h}$)
relative to the total energy of individual hadron ($M_{J/\psi}$ and $M_h$):
\begin{eqnarray}
\label{dEeq}
R_{J/\psi\textit{\rm -}h}(t) = 
\frac{G_{J/\psi\textit{\rm -}h}(t, t_{\rm src})}
{G_{J/\psi}(t, t_{\rm src})G_{h}(t+1, t_{\rm src}+1)}
\xrightarrow[t \gg t_{\rm src} ]{}  \ \ 
\exp ( -\Delta E \cdot t ),
\end{eqnarray}
where
%
%
\begin{eqnarray}
\label {dEeq2}
\Delta E = E_{J/\psi\textit{\rm -}h} -(M_{J/\psi}+M_h).
\end{eqnarray}
%
We can introduce
their relative momentum $k$  outside the interaction range through 
%
%
\begin{eqnarray}
\sqrt{M_{J/\psi}^2  + {k}^2} + \sqrt{M_h^2  + {k}^2}
= E_{J/\psi\textit{\rm -}h},
\end{eqnarray}
where $k$ should vanish as $1/L$ with increasing $L$ if there is no bound state
in this channel.

Assuming that 
the interaction range
$R$ is smaller than a half of the 
lattice size, $R < L/2$, the 
$s$-wave phase shift in a finite box, $\delta_{0}(k)$,
may be written as \cite{ML}
%
%
\begin{eqnarray}
\label{phaseAC}
\frac{2 {\cal Z}_{00}(1,q)}{L\pi^{1/2}}=k \cot \delta_{0}(k) 
 = \frac{1}{a_0} + O(k^2),
\end{eqnarray}
where $q (\equiv (\frac{kL}{2\pi})^2)$ takes a non-integer value
due to the two-particle interaction.
The function ${\cal Z}_{00}(1,q)$ is an analytic continuation of the 
generalized zeta function, ${\cal Z}_{00}(s,q)\equiv \frac{1}{\sqrt{4\pi}}\sum_{{\bf n}\in
{\bf Z}^3}({\bf n}^2-q)^{-s}$, from the region $s > 3/2$ to
$s=1$.  
The $s$-wave scattering length $a_0$ is defined through the small $k$ limit
of the above formula.

If $a_0/L$ is sufficiently small, then one can make a 
 Taylor expansion of 
 the phase-shift formula (\ref{phaseAC}) around $q^2=0$ and obtain \cite{ML}
%
%
\begin{eqnarray}
\Delta E = - \frac{2\pi a_0}{\mu L^3} 
\left( 1+ c_1 \frac{a_0}{L} + c_2 \left( \frac{a_0}{L} \right)^2 \right) +O(L^{6})
\label{largeL}
\end{eqnarray}
with $c_1=  -2.837297$ and  $c_2=  6.375183$.
$\mu$ denotes the reduced mass of two hadrons, $\mu
=M_{J/\psi}M_{h}/(M_{J/\psi}+M_{h})$.
For $\Delta E >0$, Eq.~(\ref{largeL}) with an expansion
up to $O(L^{-4})$ and that up to $O(L^{-5})$  have a real and  negative
solution for $a_0$. On the other hand, for $\Delta E <0$,
the expansion up to $O(L^{-4})$ gives no real solution for
%
%
\begin{eqnarray}
\label{dEmin}
\Delta E <  -\frac{\pi}{2|c_1|\mu L^2},
\end{eqnarray}
although the expansion
up to $O(L^{-5})$ always has a real solution.
Therefore, we can use Eq.~(\ref{dEmin}) as a necessary condition 
to test  the convergence of the large-$L$ expansion.
If this condition is not satisfied or marginally satisfied,
we need to use the full expression of (\ref{phaseAC})
to extract $a_0$ from $\Delta E$.
Eqs.~(\ref{phaseAC}-\ref{dEmin}) are
the basic formulas to be  utilized in the following sections.

\section{Numerical results and discussions}
 
We have performed simulations in quenched QCD with
the single plaquette gauge action 
%
%
\begin{eqnarray}
S_g  = \beta \sum_{p}
\left\{
1 - \frac{1}{3}{\rm Re}[{\rm Tr} U_{p}]
\right\},
\end{eqnarray}
where $\beta=6/g^2$, $U_p=U_{\mu \nu}(n)$ and $\sum_p=\sum_n\sum_{\mu<\nu}$, 
and the Wilson fermion action 
%
%
\begin{eqnarray}
S_f = \sum_{n,m} \bar{q}(n)
\Bigg\{ 
\delta_{n,m}-\kappa\sum_{\mu}
  \Big[
   (1-\gamma_\mu)U_\mu(n) \delta_{n+\mu,m}
   +(1+\gamma_\mu)U^\dagger_\mu(m) \delta_{n-\mu,m}   
   \Big]
\Bigg\}
q(m), 
\end{eqnarray}
where $\kappa$ is the hopping parameter defined as
$\kappa=\frac{1}{2(am_q+4)}$ with the lattice spacing $a$ and quark mass $m_q$.
We generate gauge ensembles at a fixed gauge coupling $\beta =6/g^2=6.2$
with three different lattice sizes, $L^3\times T =24^3 \times 48$, $32^3 \times 
48$ and $48^3 \times 48$. According to the Sommer scale \cite{Sscale1,Sscale2},
 $\beta=6.2$ in the quenched approximation
  corresponds to a lattice cutoff of $a^{-1} \simeq 2.9$ GeV,
which may be marginal to handle 
low-lying $c{\bar c}$ mesons on the lattice.
We compute the quark propagators at three values of the hopping parameter
$\kappa=\{0.1520, 0.1506, 0.1489\}$, which correspond to
 $M_{\pi}/M_{\rho}=0.68, 0.83, 0.90$.
$\kappa_c=0.1360$ is reserved for the charm-quark mass.
Simulation parameters and the number of the gauge samples are summarized 
in Table \ref{lConst}.
Preliminary accounts for $L=$24 and  32 cases are given in Ref.~\cite{YSHH}.

For the update algorithm, we adopt the Metropolis algorithm with 20 hits at each
link update. The first 10000 sweeps are discarded for thermalization.
The gauge ensembles in each simulation are separated by 1200 ($L=48$),
800 ($L=32$) and 600 ($L=24$) sweeps. For the matrix inversion, we use
BiCGStab algorithm and adopt the convergence condition $|r|<10^{-8}$ for the residues.
We calculate smear-point quark propagators with a box-type smeared source
which are located at $t_{\rm src}=6$ for the light quarks and at $t_{\rm src}=5$
for the charm quark. To enhance the coupling to ground states of hadrons, 
 we choose a $24^3$ box source, of which spatial lattice size is about 1.6 fm, on all
three volumes, 
instead of a wall ($L^3$ box).
 Since the box-type smeared 
source is gauge variant, Coulomb gauge fixing is carried out by
a combination of an $SU(2)$ subgroup method and an overrelaxed steepest descent method.
Details of this procedure may be found in Ref.~\cite{gFix}.
To perform the precise parity projection for the nucleon, we adopt a procedure to take 
an average of two quark propagators which are subject to periodic and antiperiodic boundary
conditions in time as described in Refs.~\cite{{SBO},{KSSS}}.

\subsection{Energy shift $\Delta E$}

 Hadron masses computed with the conventional 
single exponential fit  are summarized in Table~\ref{table-mass}. 
 $\kappa_c=0.1360$ reproduces the mass of $J/\psi(3097)$ approximately.
 The mass of pseudoscalar meson becomes
$M_{\pi} \approx$ 0.6 GeV for $\kappa=0.1520$, 0.9 GeV for $\kappa=0.1506$
and 1.2 GeV for $\kappa=0.1489$ in our simulations.
 
The energy shifts of two-hadron systems 
($J/\psi$-$\pi$, $J/\psi$-$\rho$ and $J/\psi$-$N$) 
 are determined 
 by the large-$t$ behavior of the ratio $R_{J/\psi \textit{\rm -}h}$ 
defined in Eq.~(\ref{dEeq}). To find appropriate temporal windows for fitting,
 we define the effective energy shift by
\begin{eqnarray} \label{defMeff}
\Delta E_{\rm eff}(t) = \ln \frac{R_{J/\psi \textit{\rm -}h}(t)}
{R_{J/\psi \textit{\rm -}h}(t+1)},
\end{eqnarray}
which should show a plateau for large Euclidean time ($t \gg t_{\rm src}$)
 and approaches to the true energy shift $\Delta E$, 
 if the statistics is sufficiently large.

\begin{figure}[h]
\begin{center}
\includegraphics[scale=0.33]{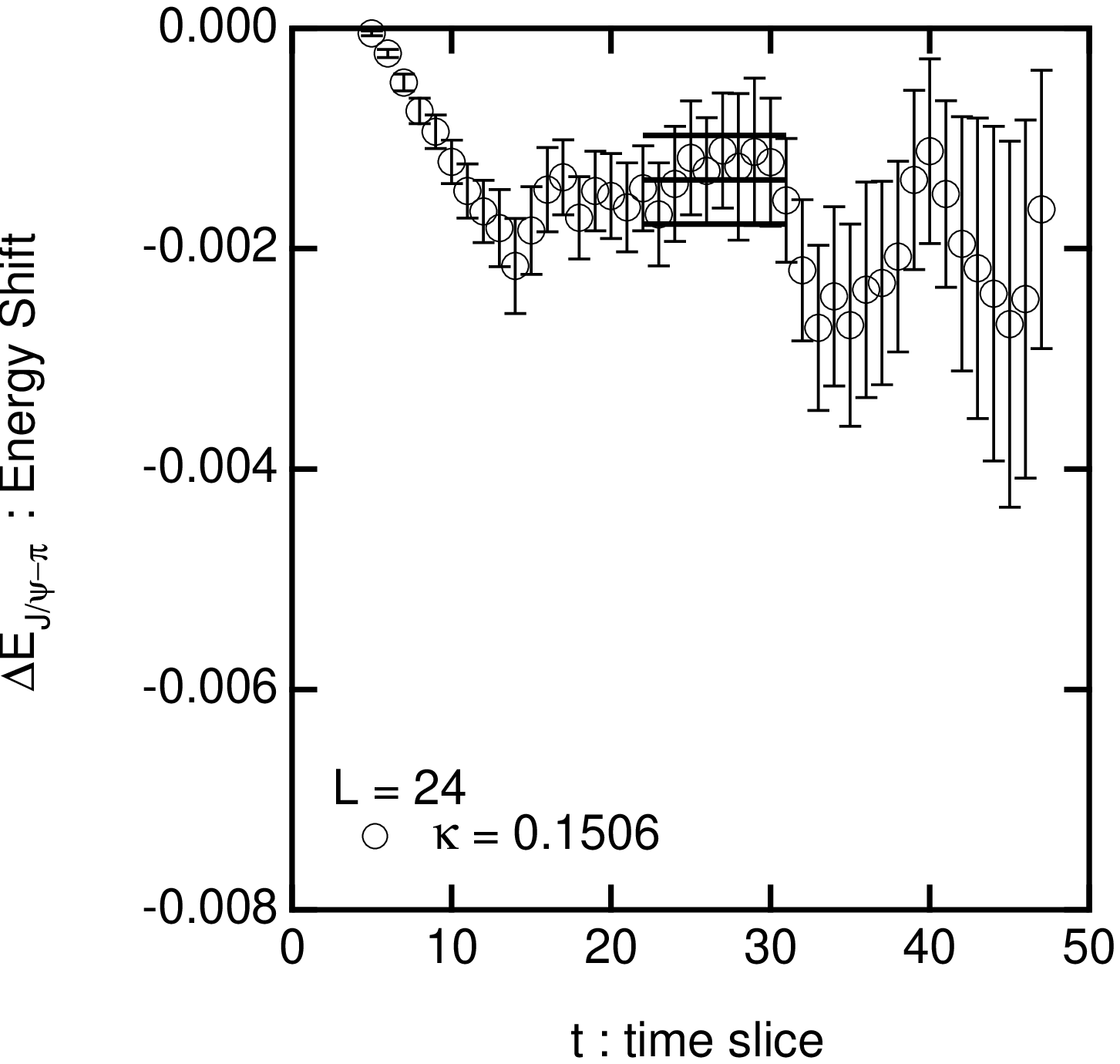}
\includegraphics[scale=0.33]{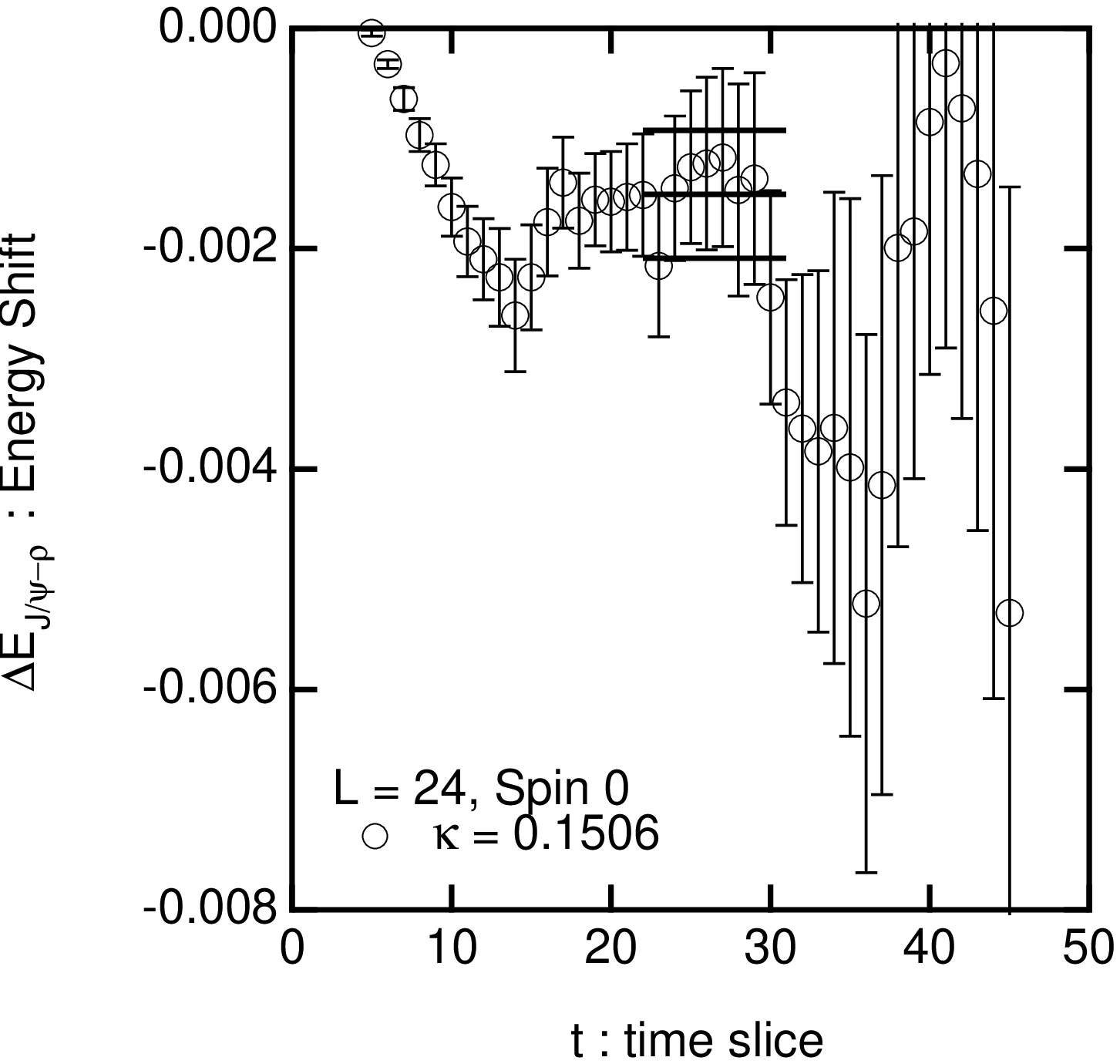}
\includegraphics[scale=0.33]{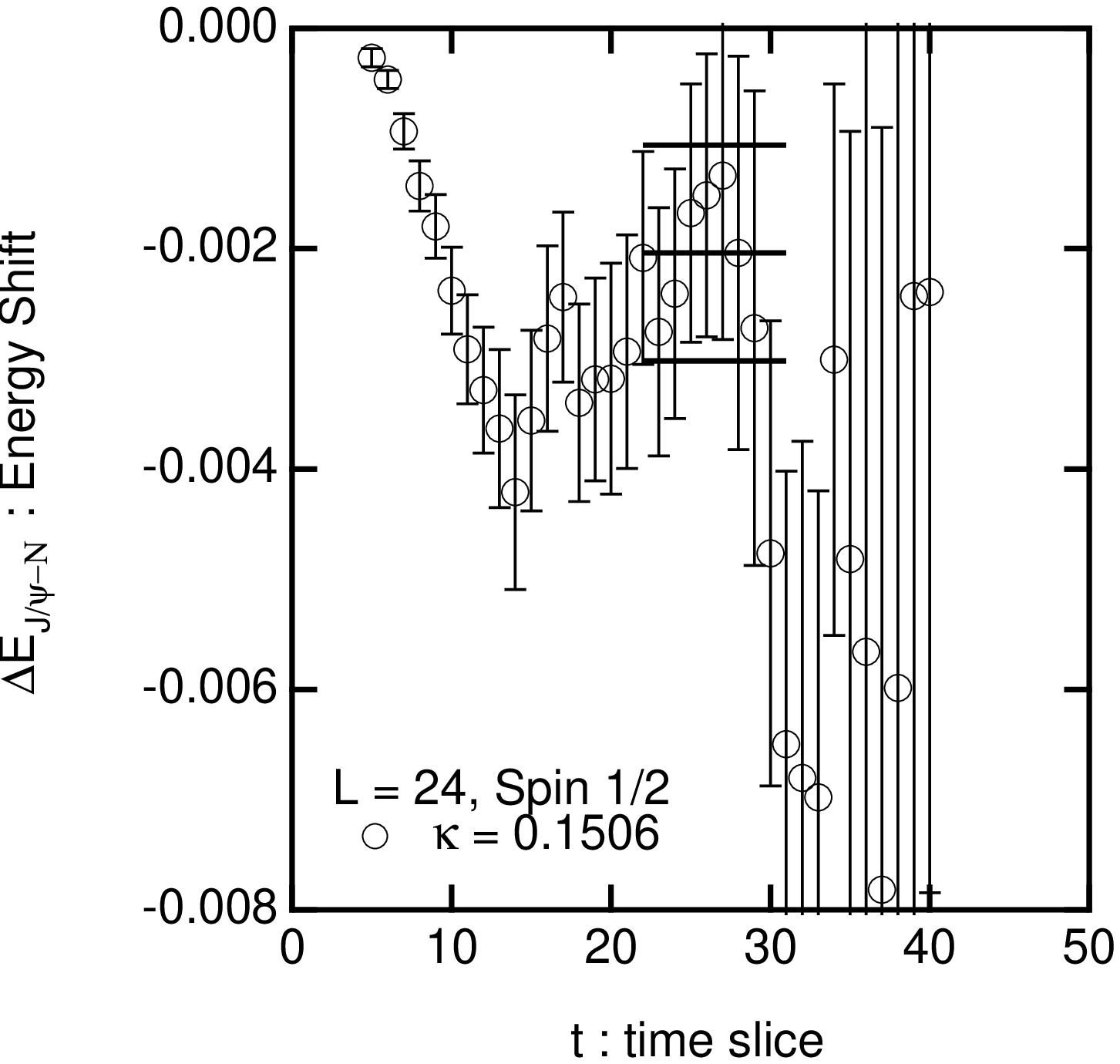}
\end{center}
\vspace{.5cm}
\begin{center}
\includegraphics[scale=0.33]{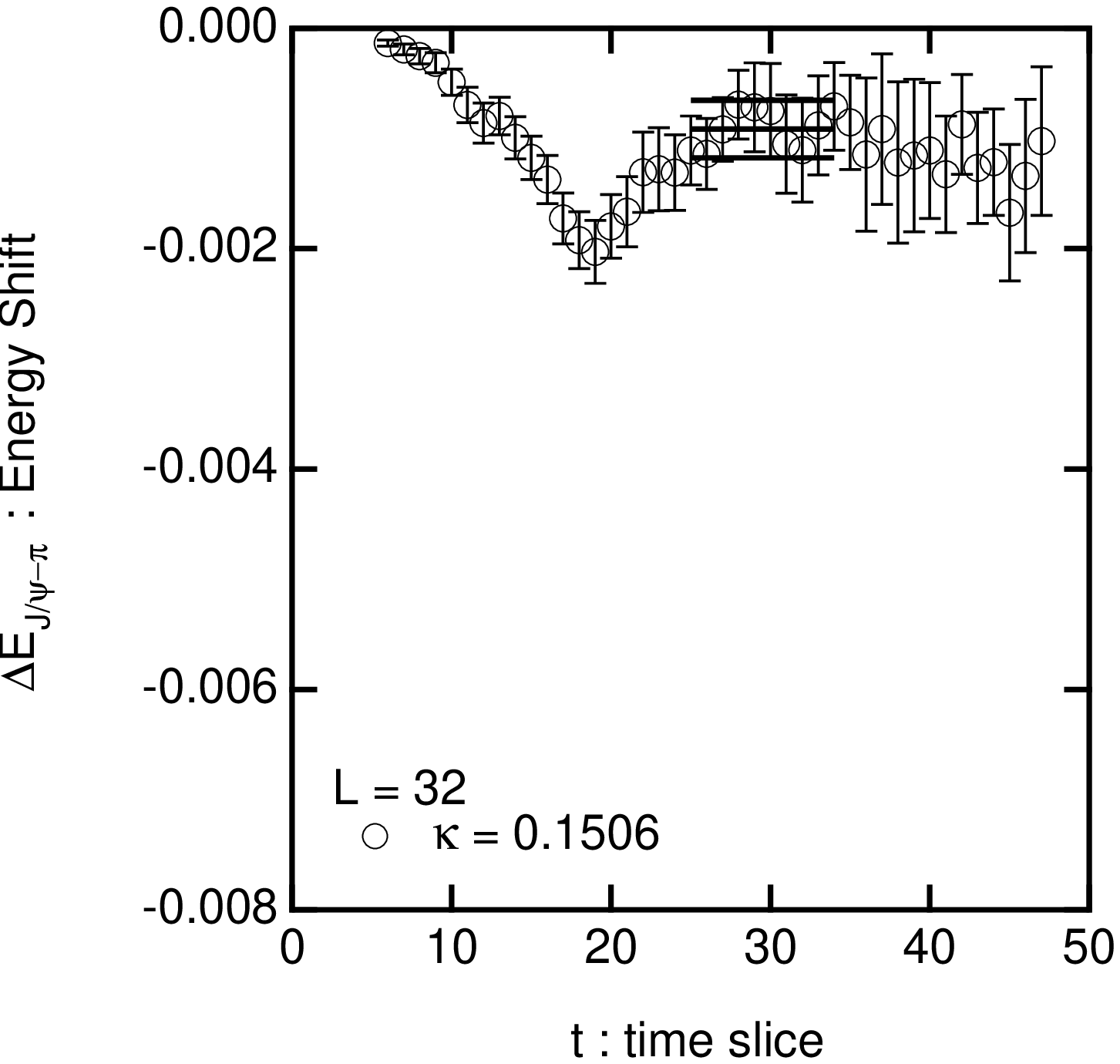}
\includegraphics[scale=0.33]{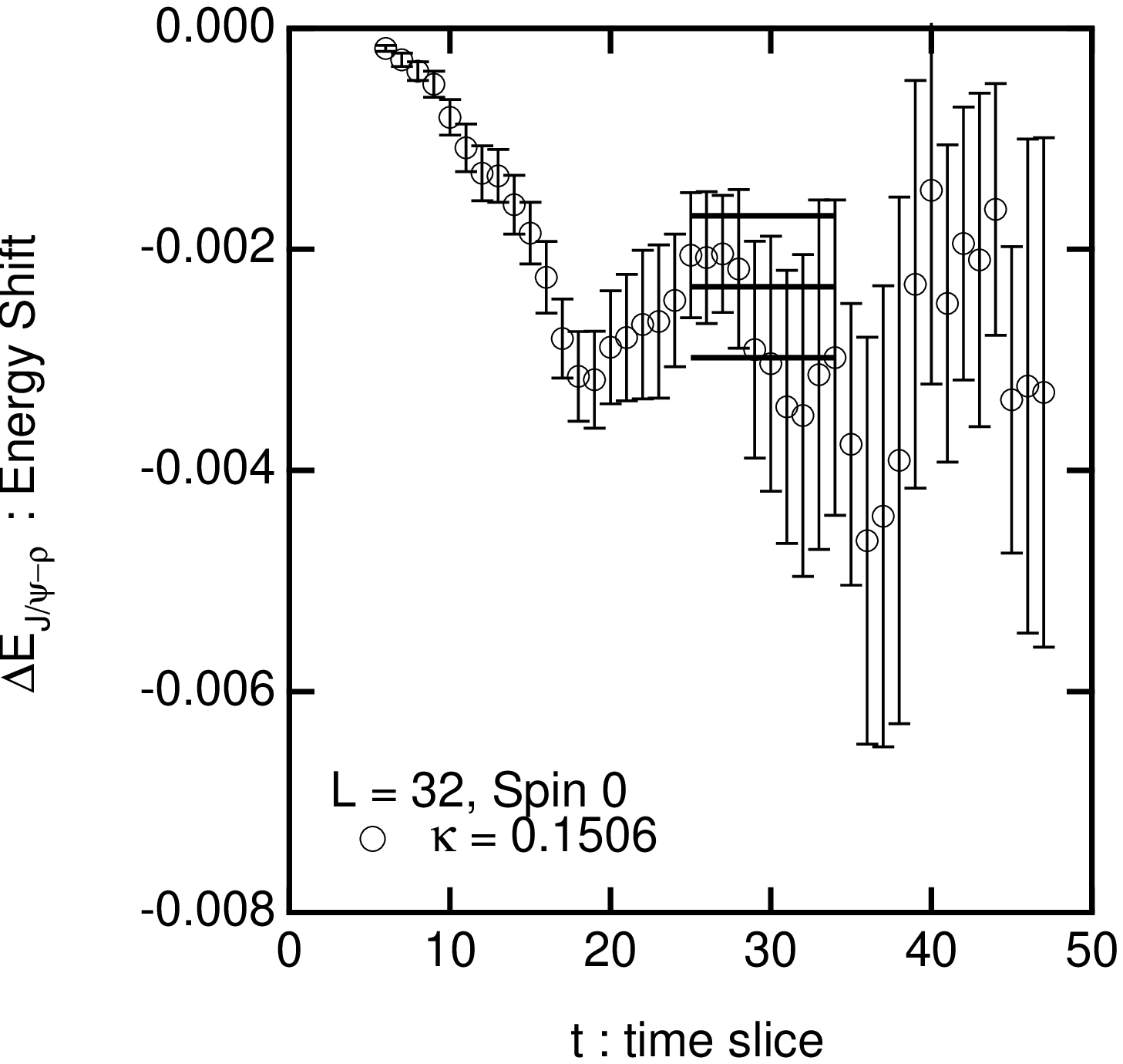}
\includegraphics[scale=0.33]{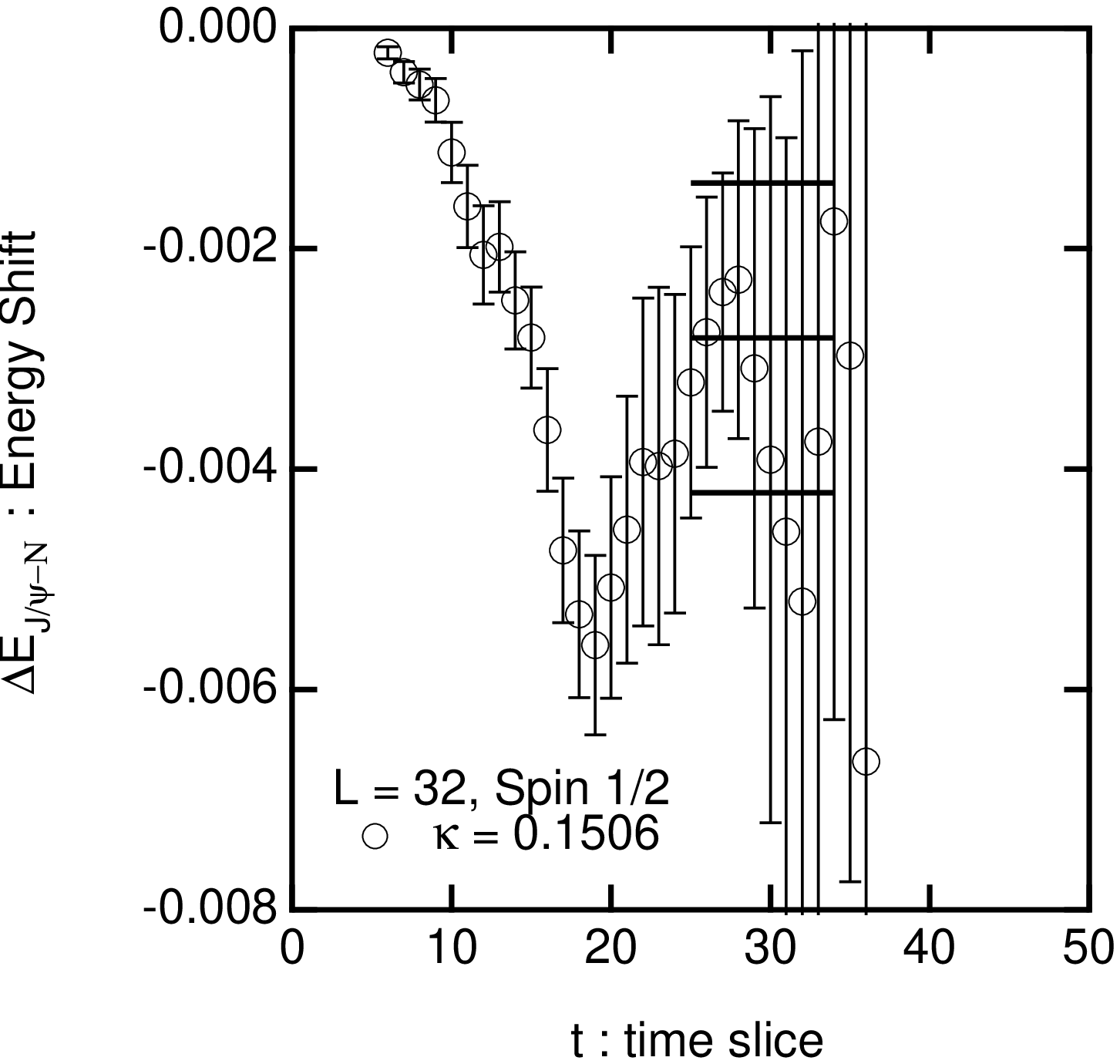}
\end{center}
\vspace{.5cm}
\begin{center}
\includegraphics[scale=0.33]{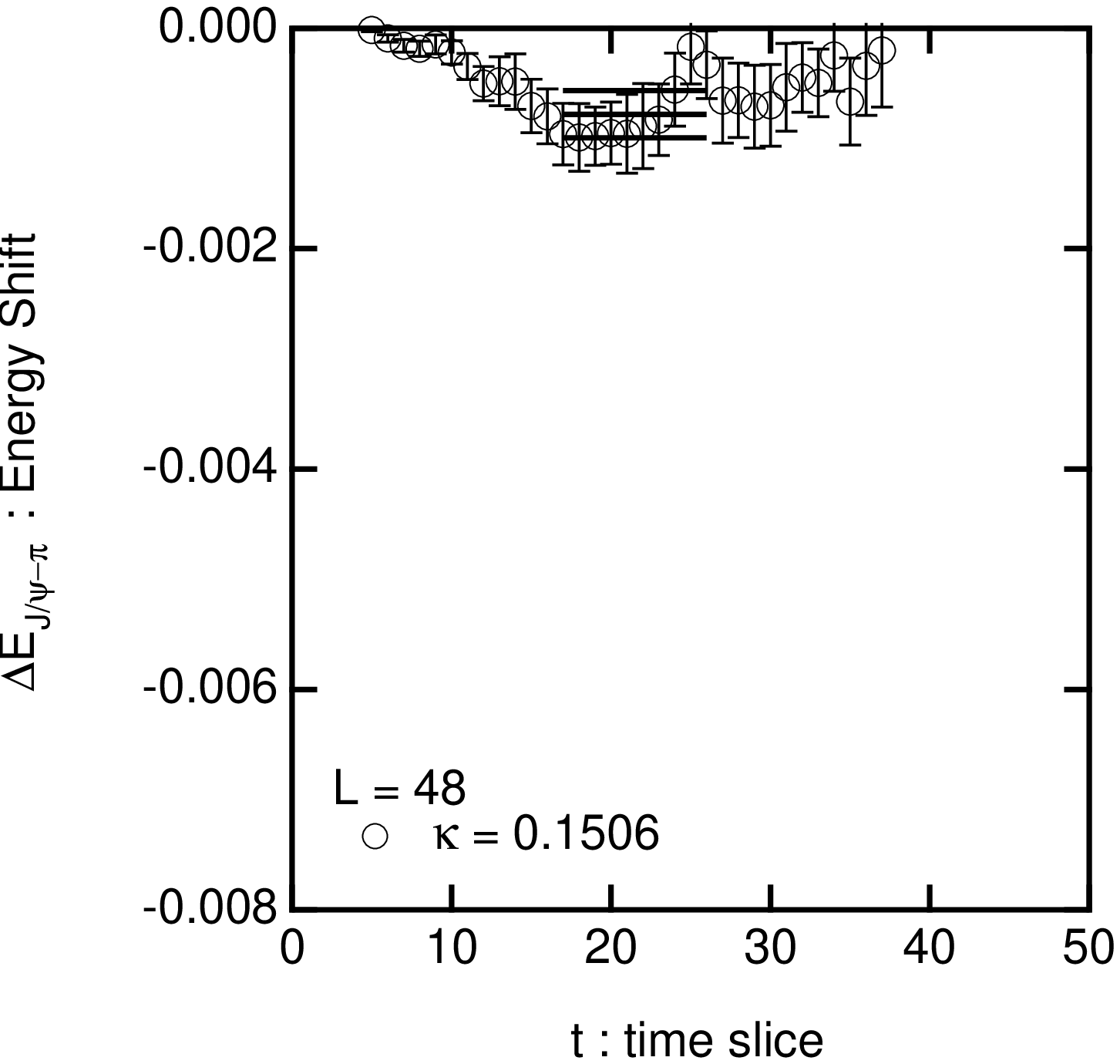}
\includegraphics[scale=0.33]{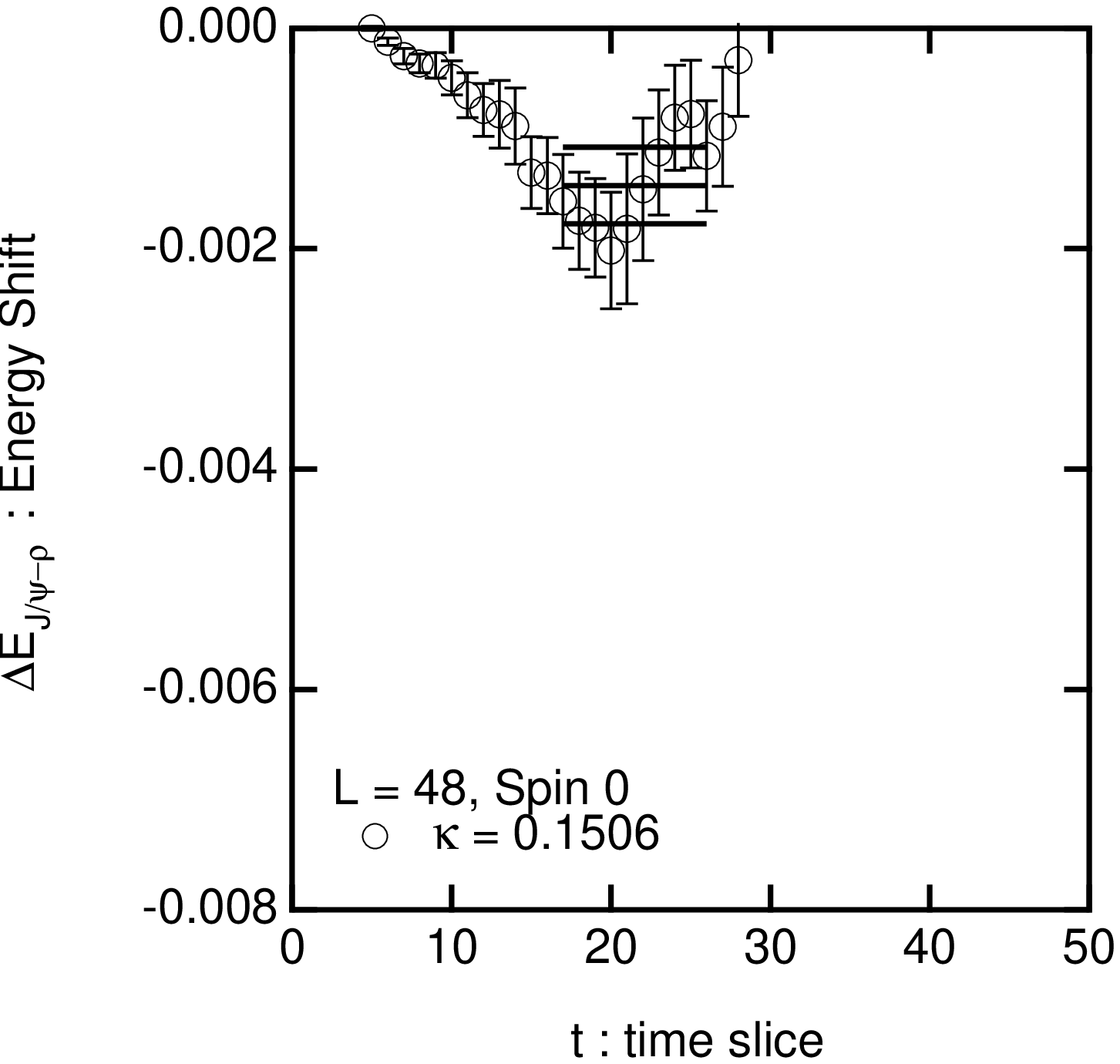}
\includegraphics[scale=0.33]{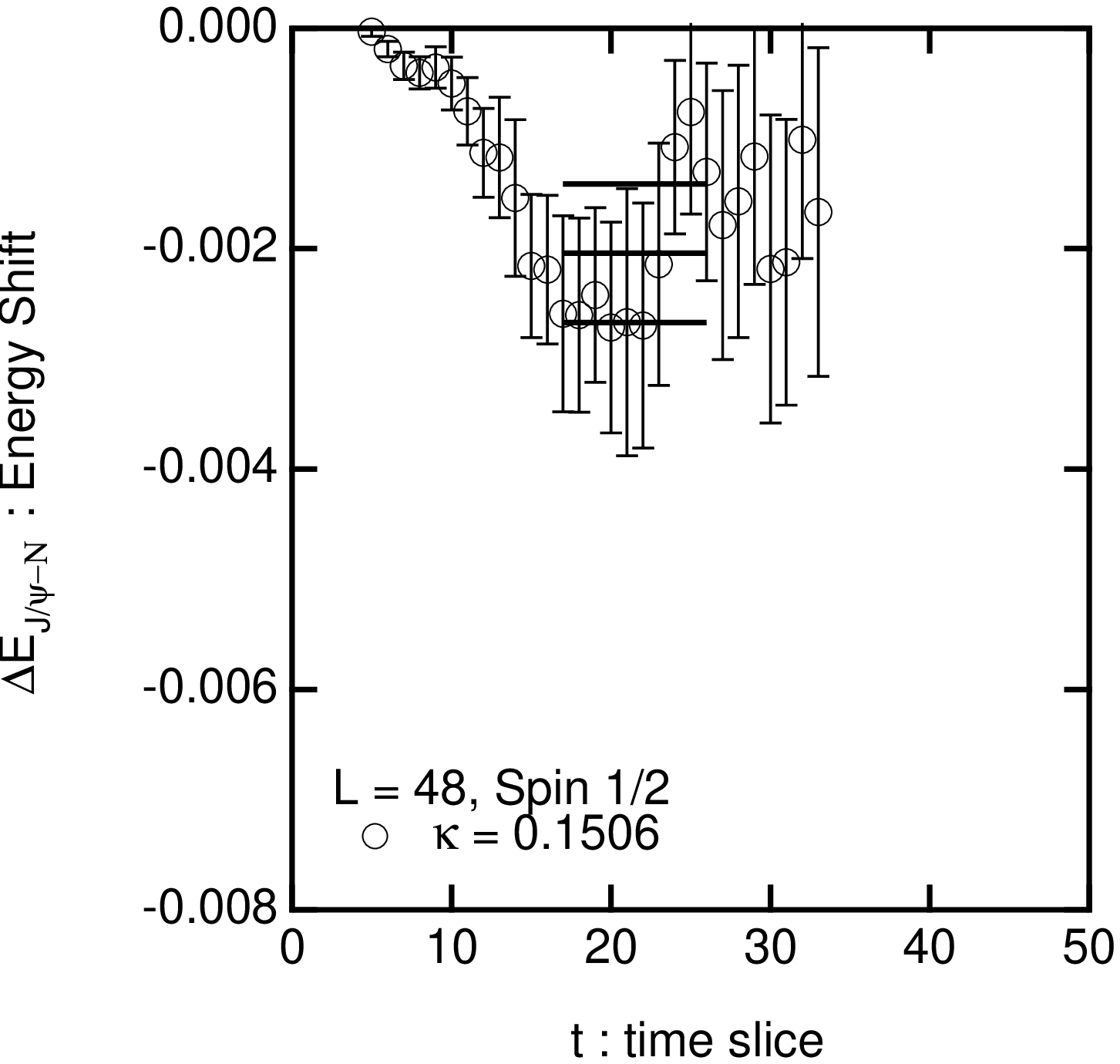}
\caption{
The effective energy shifts $\Delta E$ in lattice units as a function of the time slice $t$. 
Hopping parameters of light hadrons and the $J/\psi$
are fixed to be $\kappa = $ 0.1506 and 0.1360, respectively.
Top (middle, bottom) panels are for $L=$24 (32, 48).
Left (middle, right) panels are for the 
$J/\psi$-$\pi$ ($J/\psi$-$\rho$ in spin-0, $J/\psi$-$N$ in spin-1/2) channels.
Source locations of light hadrons and the $J/\psi$ are at $t_{\rm src}=6$ and 5, respectively.
}\label{Meff_dE}
\end{center}
\end{figure}

As typical examples of our simulations, in Fig.~\ref{Meff_dE} we show the
  effective mass $\Delta E_{\rm eff}(t)$ as a function of $t$ 
   for $L=24$ (top panels), 32 (middle panels) and 48 (bottom panels) at $\kappa=0.1506$. 
In the figure, the left panels are for the $J/\psi$-$\pi$ channel, 
the middle ones for the spin-0 $J/\psi$-$\rho$ channel,
 and the right ones for the spin-1/2 $J/\psi$-$N$ channel.
Statistical uncertainty of $\Delta E_{\rm eff}(t)$ is estimated by a single elimination 
jack-knife method. 

As seen from this figure, the effective energy shift is quite small compared to the mass of each 
hadron. Nevertheless, strong correlation between the numerator and denominator 
in the ratio $R_{J/\psi \textit{\rm -}h}$ can help to expose such a tiny shift.
  An important observation is that 
 the negative energy shift is found in all channels of Fig.~\ref{Meff_dE}, 
  which implies that there exist attractive
 interactions between $J/\psi$ and light hadrons ($\pi$, $\rho$ and $N$).

One should, however, notice that
  the signals become quite noisy for $t$ far away from the sources at
  $t_{\rm src}=5$ and 6. Therefore, we need to make 
   appropriate choice of the fitting window in $[t_{\rm min}, t_{\rm max}]$
    to extract
    $\Delta E$ from the data.  
 Since the two-point correlation functions 
 are well dominated by ground state hadrons ($J/\psi$, $\pi$, $\rho$ and $N$)
 for  $t > 20$, a reasonable choice of the lower bound is
  $t_{\rm min} \sim 20$. The upper bound may be chosen
  so that we have a reasonable value of $\chi^2/{\rm dof}$ in the fitting 
   interval.  Three horizontal solid lines in
  each panel of Fig.~\ref{Meff_dE} 
  represent the fitting range thus determined together
   with the value of $\Delta E$  and its 1$\sigma$
 deviation obtained  by a covariant single exponential fit
 to the ratio $R_{J/\psi \textit{\rm -}h}$. 
The fitting range is chosen to be the same for the data with equal $L$.
  The results  of energy shift obtained in the above procedure
are summarized  in Tables~\ref{Table:EsftL24}, \ref{Table:EsftL32}
and \ref{Table:EsftL48}.

\subsection{Quark mass dependence of the energy shifts}

\begin{figure}[htbp]
\begin{center}
\includegraphics[scale=0.33]{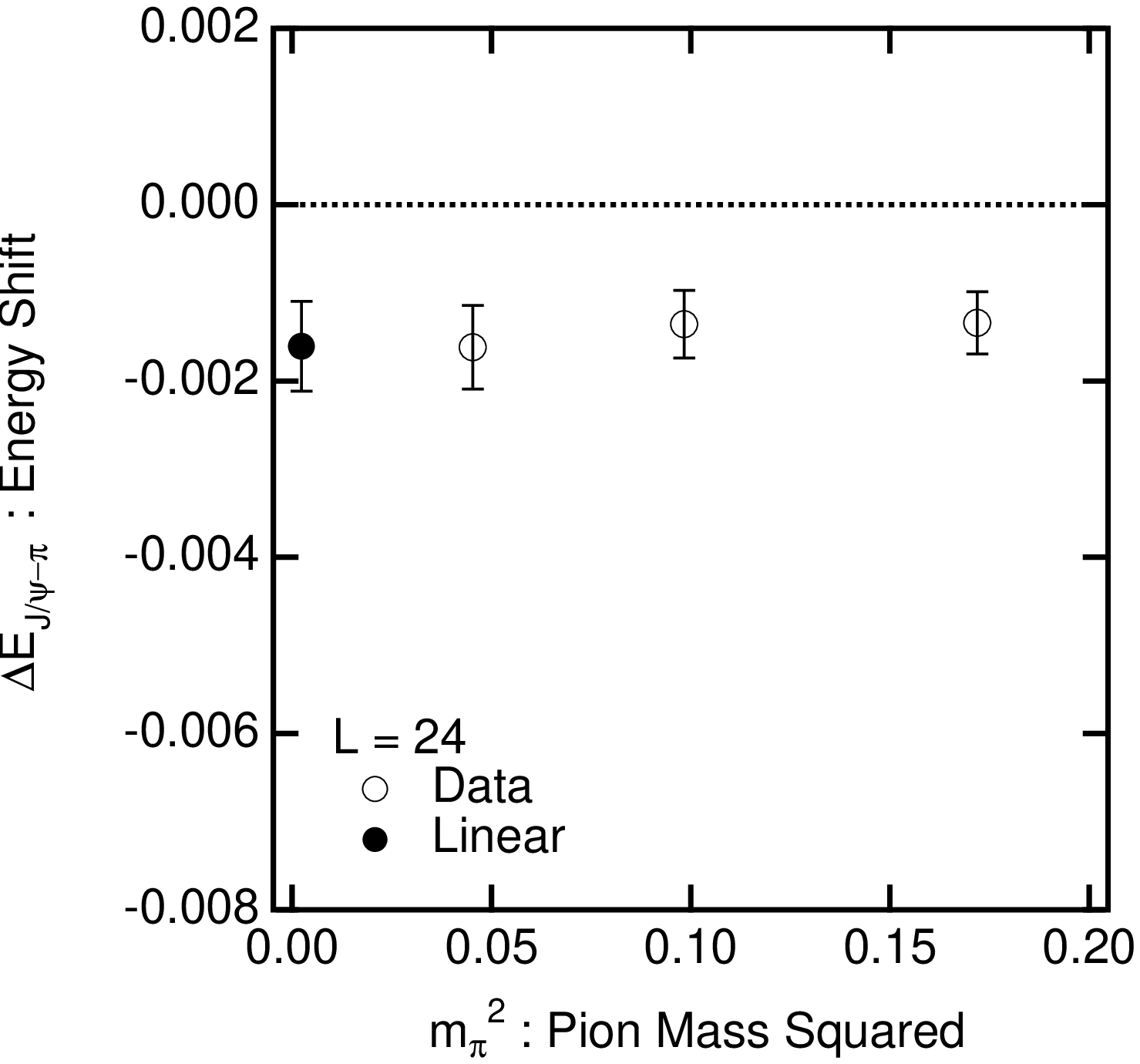}
\includegraphics[scale=0.33]{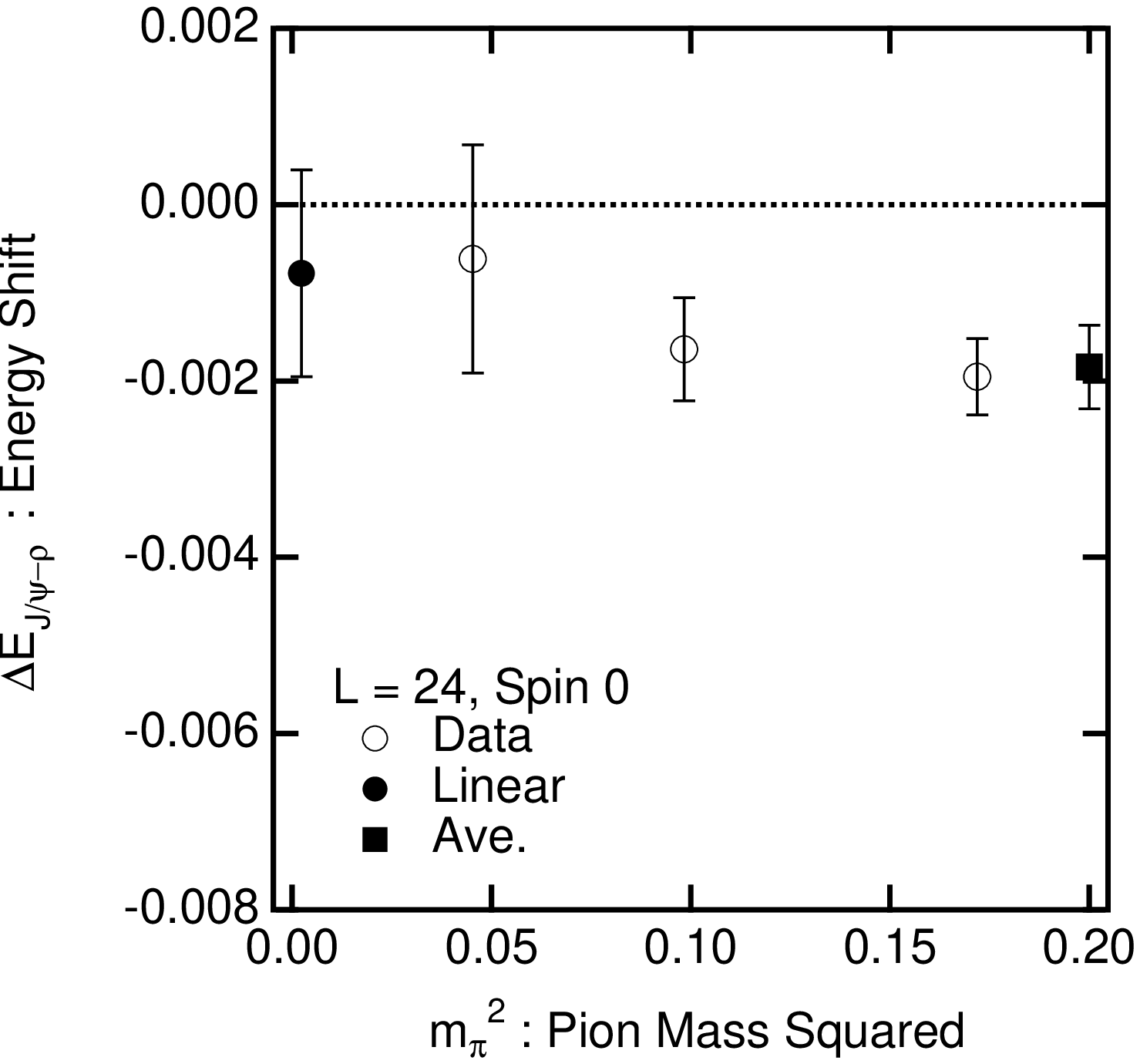}
\includegraphics[scale=0.33]{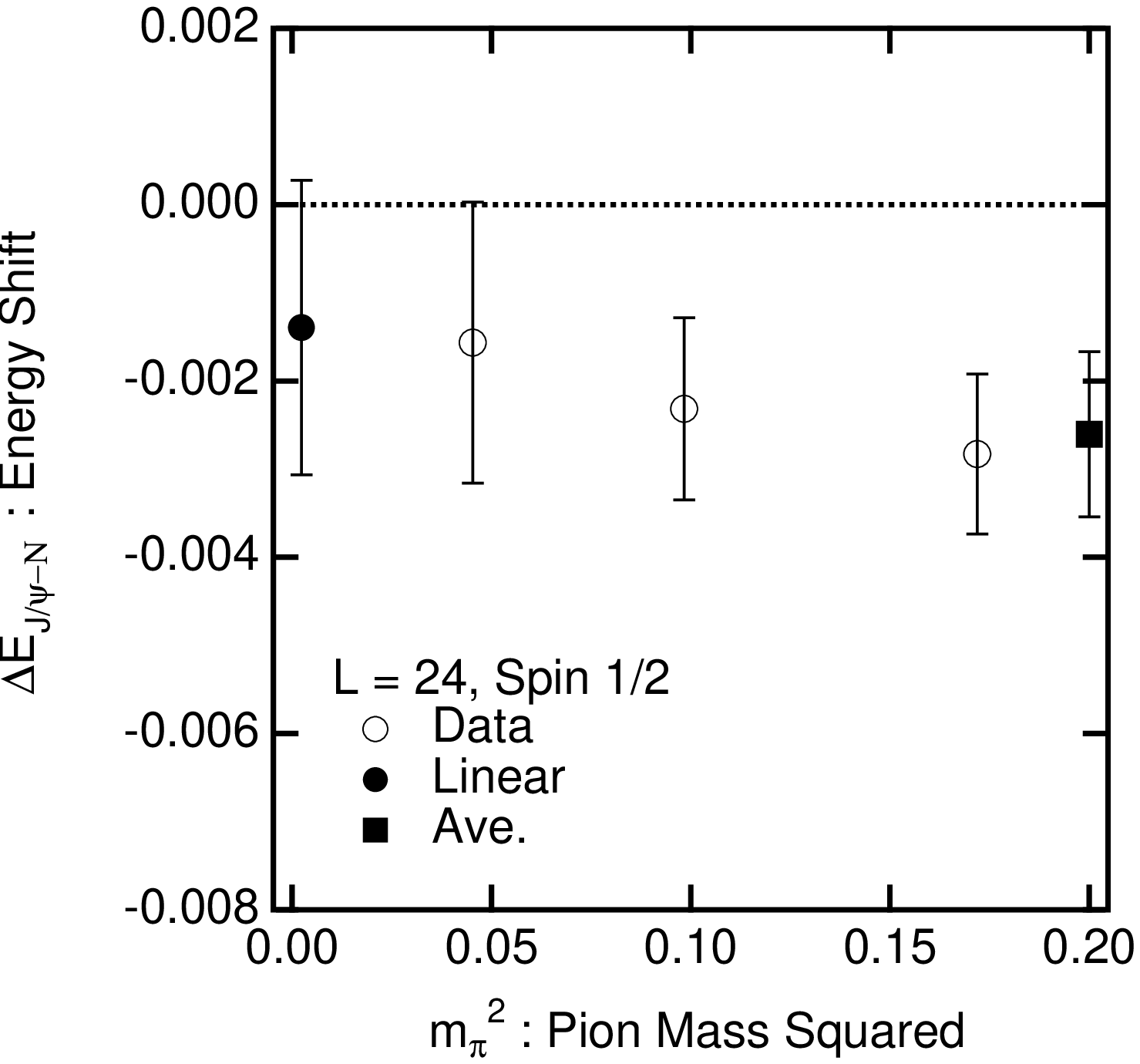}
\end{center}
\vspace{.5cm}
\begin{center}
\includegraphics[scale=0.33]{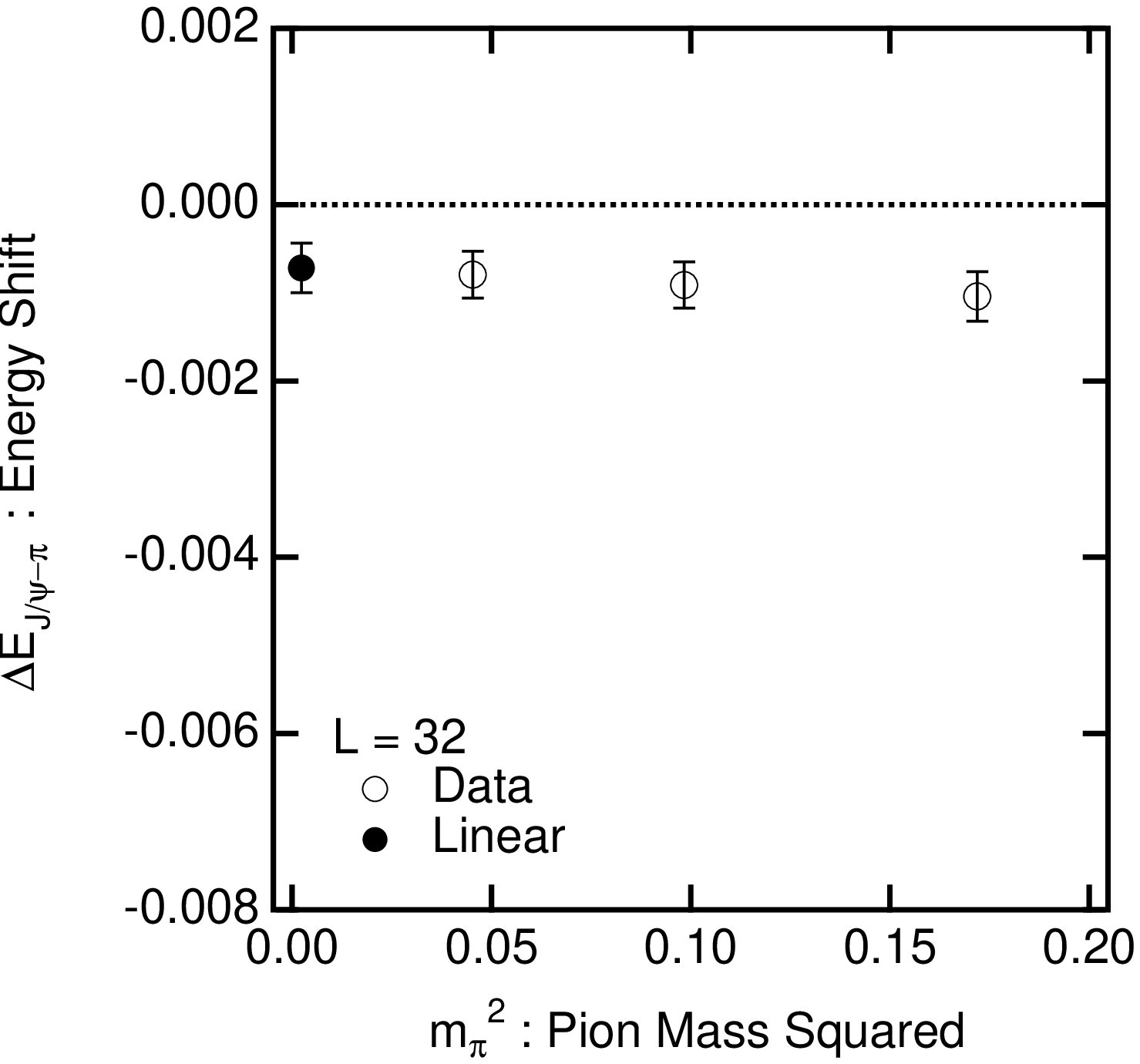}
\includegraphics[scale=0.33]{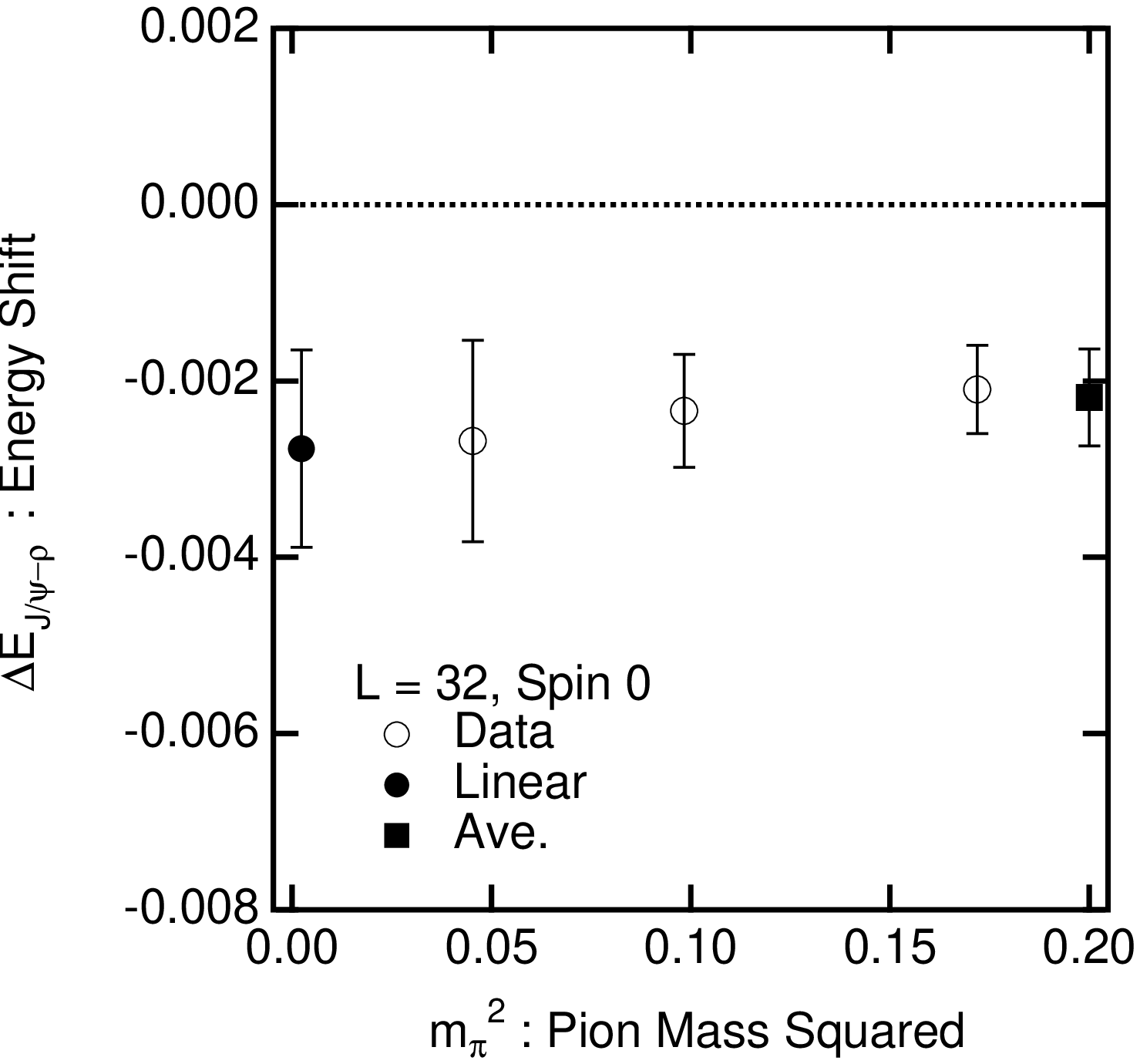}
\includegraphics[scale=0.33]{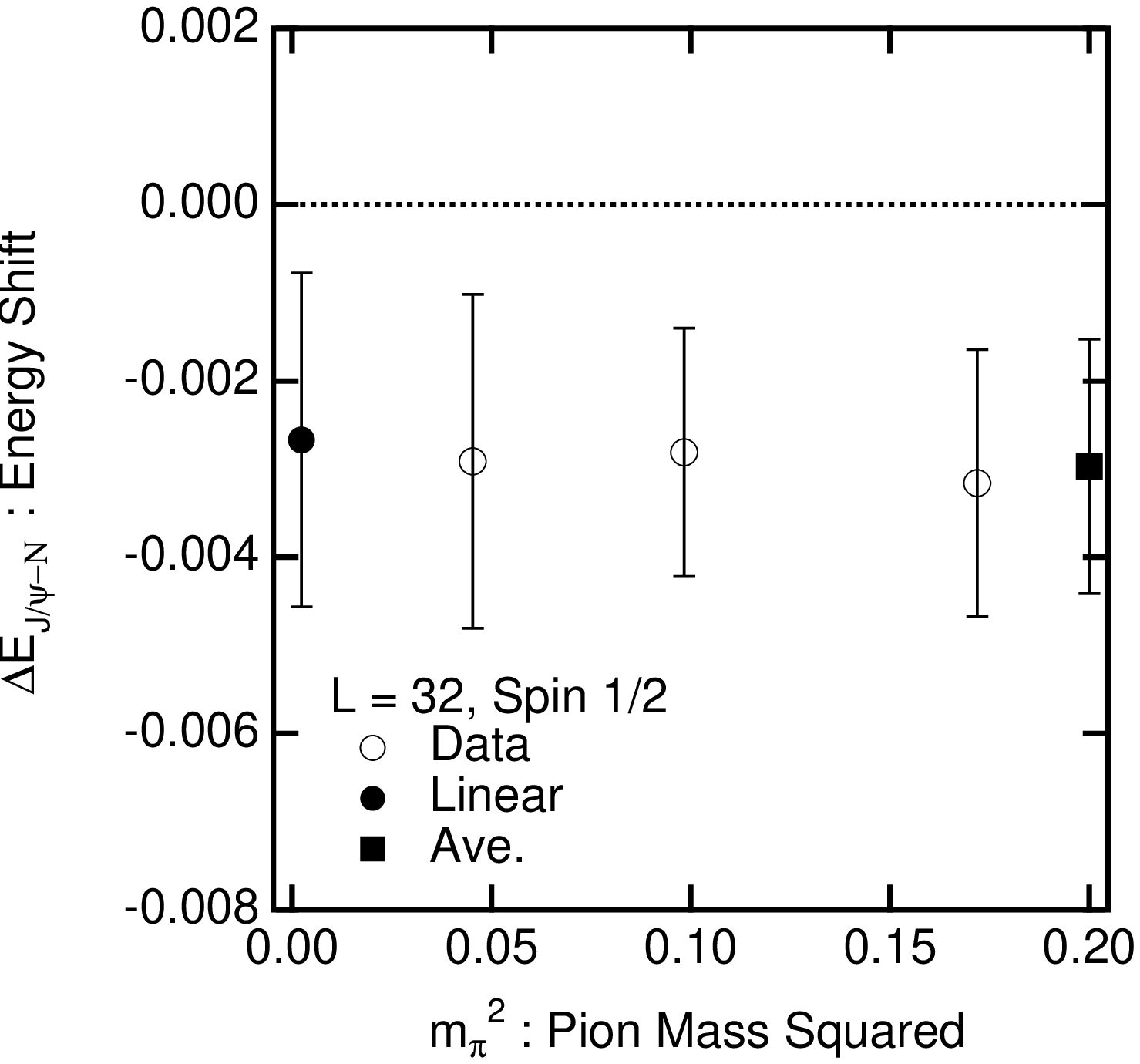}
\end{center}
\vspace{.5cm}
\begin{center}
\includegraphics[scale=0.33]{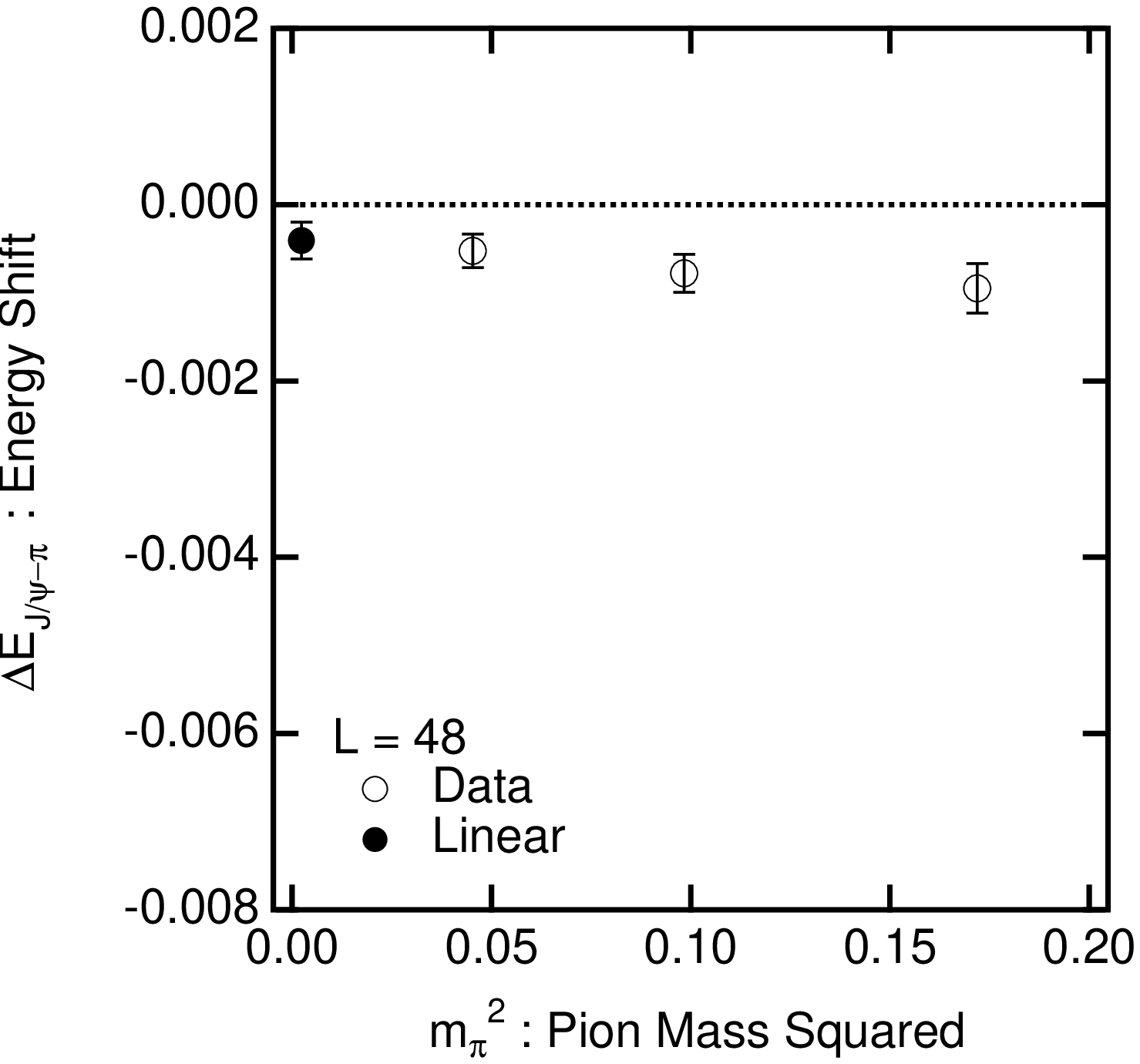}
\includegraphics[scale=0.33]{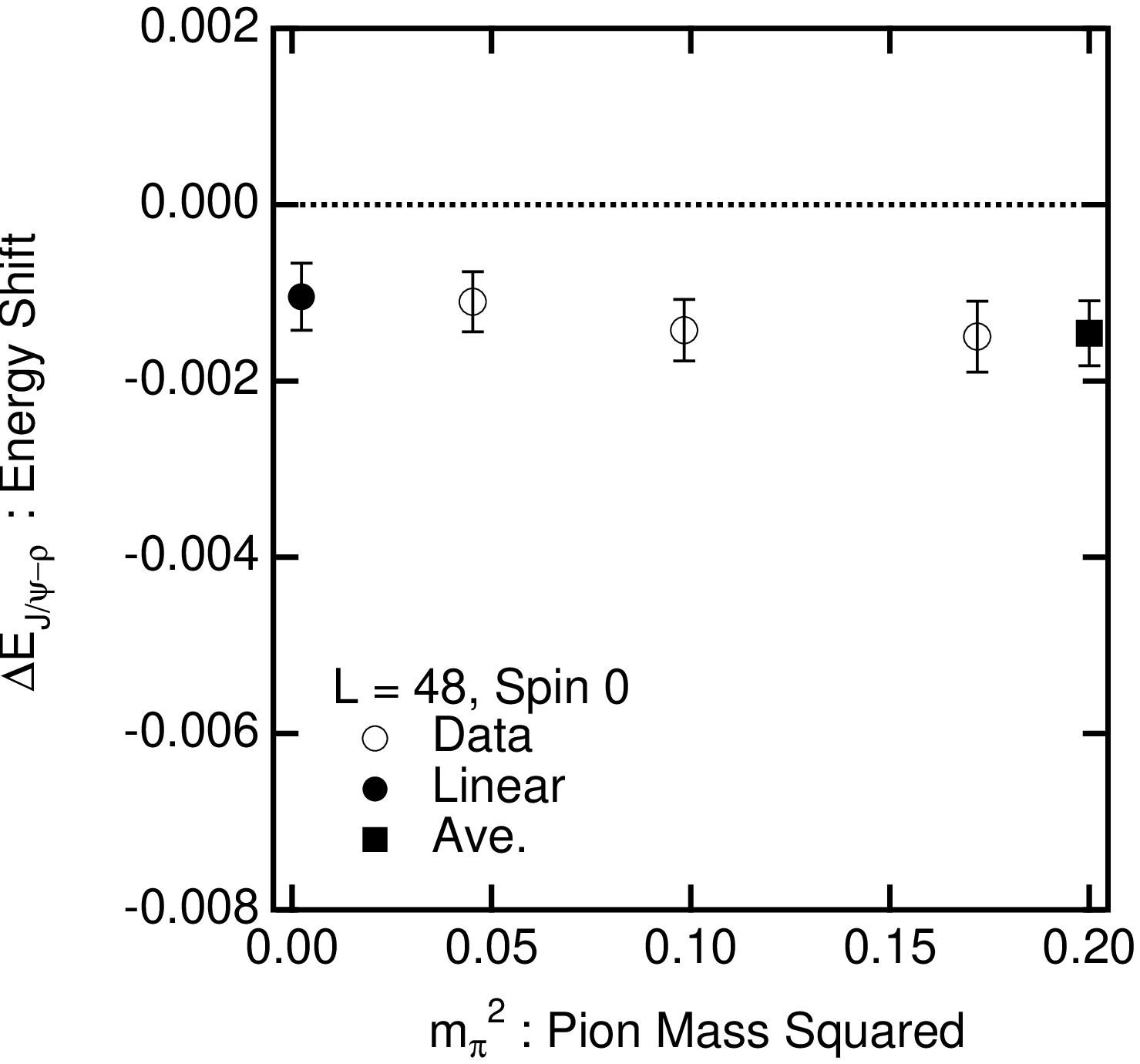}
\includegraphics[scale=0.33]{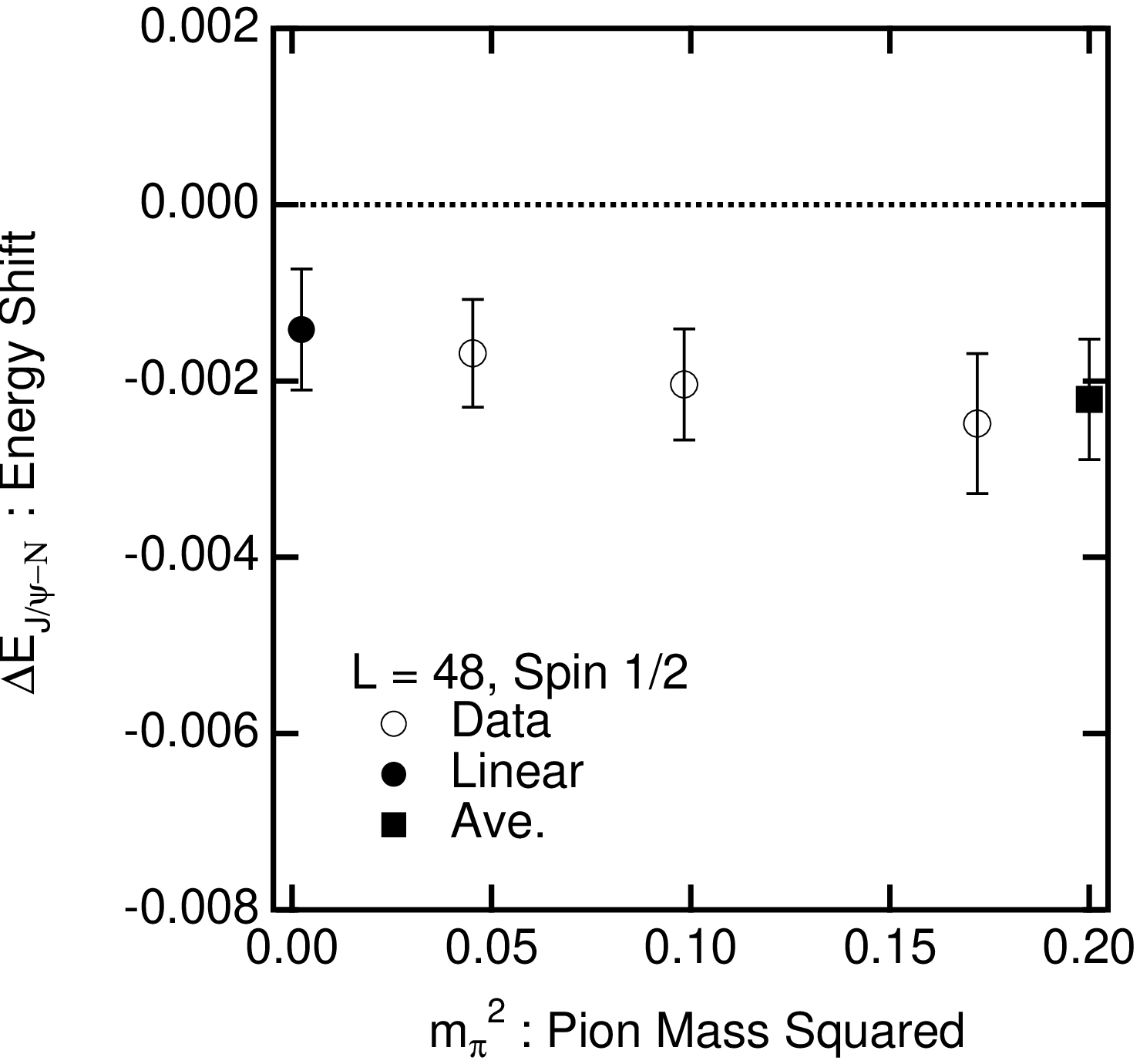}
\caption{
The energy shifts $\Delta E$ as a function of the squared pion mass $M_\pi^2$ in lattice units.
Upper (middle, lower) panels are for $L=24$ (32, 48).
Left (middle, right) panels are for the $J/\psi$-$\pi$ ($J/\psi$-$\rho$ in spin-0, 
$J/\psi$-$N$ in spin-1/2) channels. 
Open circles are the values fitted by the criterion obtained from the four-point function  
at the hopping parameters of light hadrons ($\kappa =$ 0.1489, 0.1506 and 0.1520).
Full circles are the values extrapolated to the physical point ($M_\pi=$140 MeV)
by using Eq.~(\ref{chiext}).
Full squares are the values from weighted averages of heavy-quark results 
($\kappa =$ 0.1489 and 0.1506).
}\label{dEmDep}
\end{center}
\end{figure}

Fig.~\ref{dEmDep} shows the quark mass dependences of the energy shift 
$\Delta E$ for $L=24$ (upper panels), $L=32$ (middle panels) and 
$L=48$ (lower panels). 
The left panels are for the $J/\psi$-$\pi$ channel, 
the middle ones for the spin-0 $J/\psi$-$\rho$ channel and
 the left ones for the spin-1/2 $J/\psi$-$N$ channel.
 Open circles show the results evaluated at three values of the 
 hopping parameter for the light hadrons, $\kappa=0.1520$, 0.1506 and 0.1489.
 We do not find appreciable spin dependence of the energy shift
 within the error bars in the $J/\psi$-$\rho$ 
 and $J/\psi$-$N$ channels.
 
The $J/\psi$-$\pi$ channel has a special feature in the sense that 
 the low-energy pion decouples from any hadrons 
  in the chiral limit as discussed in Sec.~1.
  In other words, $\Delta E$ of the  $J/\psi$-$\pi$ system 
  should vanish in simultaneous chiral and infinite volume limits.
  The top, middle and bottom panels for the 
  $J/\psi$-$\pi$ channel in Fig.~\ref{dEmDep} indeed suggest
   such a tendency. This will be quantified later in Sec.~\ref{subsec:vol-dep-a}.

To make an extrapolation of  $\Delta E$ to the chiral limit, we adopt  
a simple formula:
\begin{eqnarray}
a\Delta E= c + c' (aM_\pi)^2,
\label{chiext}
\end{eqnarray}
where $c$ and $c'$ are determined numerically from the data.
 In Fig.~\ref{dEmDep}, full circles represent
the energy shifts linearly extrapolated by
Eq.~(\ref{chiext}) with the physical pion mass squared ($M_\pi=$140 MeV). 
 For the $J/\psi$-$\rho$ and $J/\psi$-$N$ channels, 
 we show also the results (full squares)
 obtained from the weighted average of the data for two heavy-quark masses 
 ($\kappa=0.1506$ and 0.1489).
  The purpose is to check the sensitivity from the lightest data points
 which have relatively large error bars.
 We find that results of full circles and squares 
 are in agreement with each other within the errors,
 and hence we quote the results evaluated by Eq.~(\ref{chiext}) in the following sections.

\subsection{Volume dependence of the energy shifts}
\label{sec:volume-dE}

\begin{figure}[htbp]
\begin{center}
\includegraphics[scale=0.33]{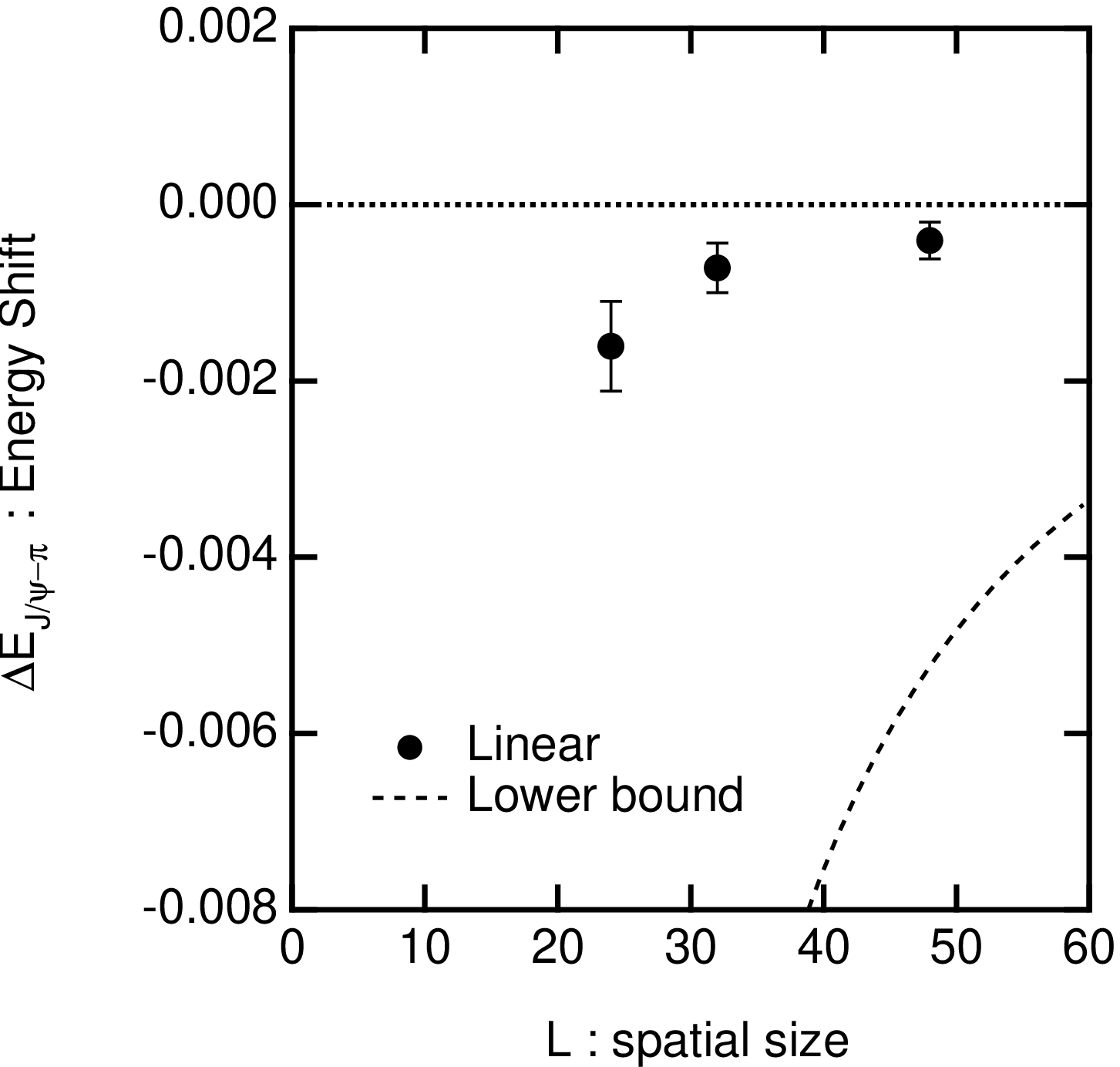}
\includegraphics[scale=0.33]{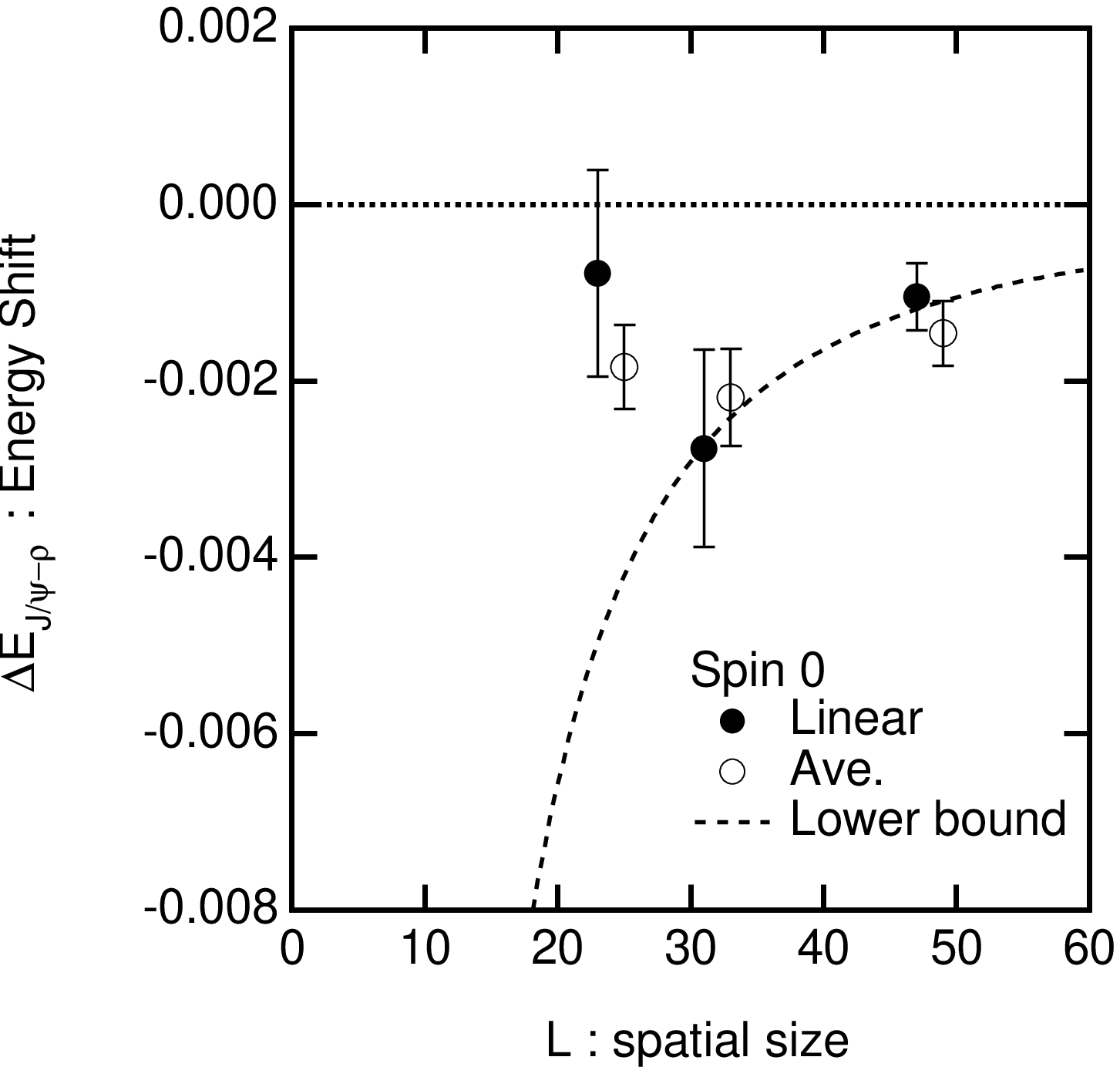}
\includegraphics[scale=0.33]{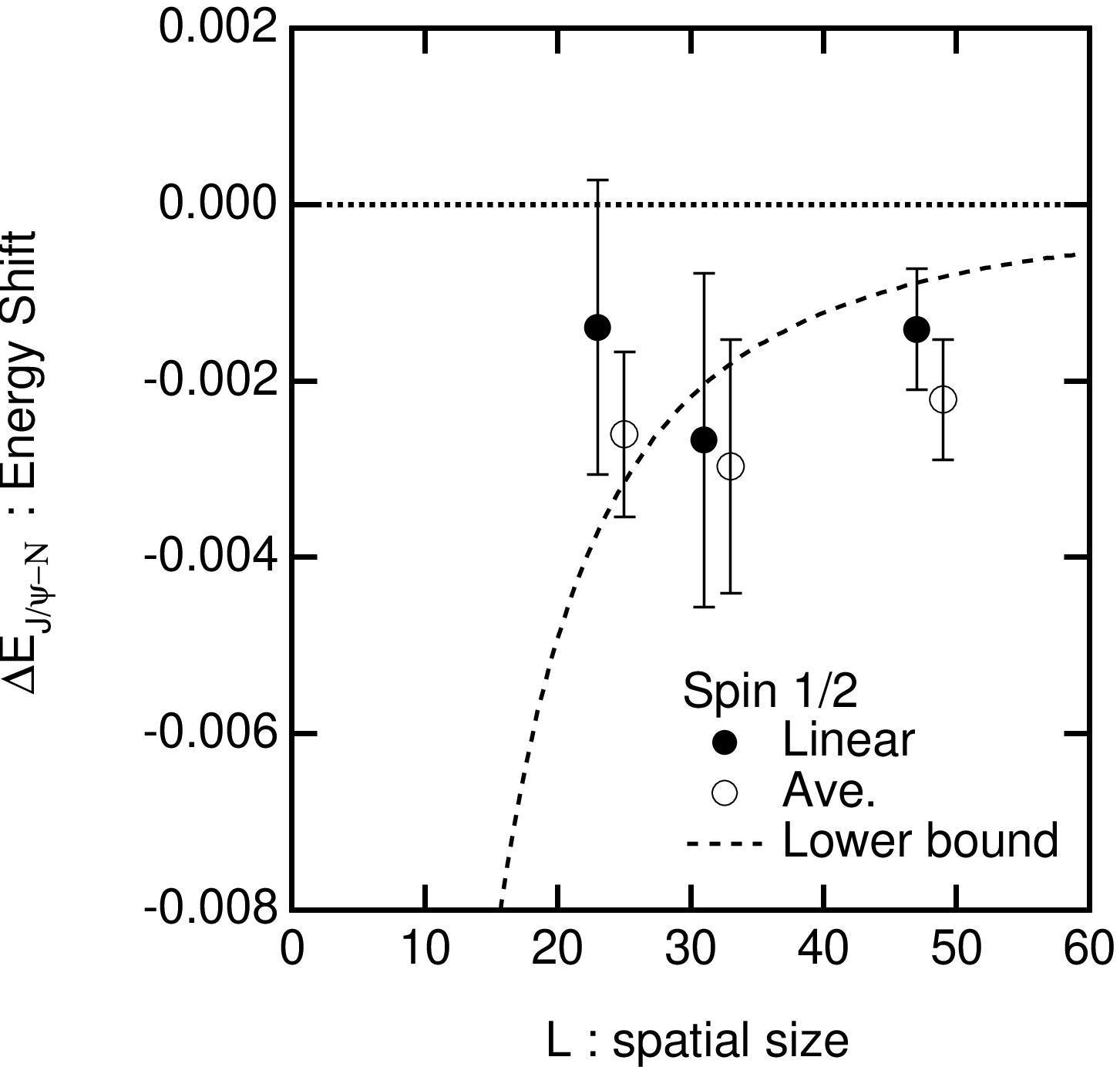}
\caption{
The energy shifts $\Delta E$ as a function of spatial size $L$ in lattice units.
Left (middle, right) panel is for the $J/\psi$-$\pi$ ($J/\psi$-$\rho$ in spin-0, 
$J/\psi$-$N$ in spin-1/2) channel. 
Full and open circles represent the values obtained from the linear chiral-extrapolation
to the physical point, and the weighted average, respectively. 
The dashed curves show the lower boundary for the convergence of the large-$L$ expansion
of $\Delta E$.
}\label{dEVDep}
\end{center}
\end{figure}

In Fig.~\ref{dEVDep},
we show the volume dependence of the energy shift $\Delta E$ at the physical
point ($M_{\pi}$=140 MeV).
The left panel is for the $J/\psi$-$\pi$ channel, 
the middle for the spin-0  $J/\psi$-$\rho$ channel
 and the right for the spin-1/2 $J/\psi$-$N$ channel.
  The horizontal axis denotes the spatial size $L$ and the vertical axis  
the energy shift $\Delta E$.
   The dashed lines indicate the lower boundary given 
  in Eq.~(\ref{dEmin}): If the data points come below the dashed
   lines, the $1/L$-expansion of the phase-shift formula is no longer
   justified.

The volume dependence of the energy shift $\Delta E$ 
is a key quantity to discuss the validity of the large-$L$ expansion of the 
phase-shift formula. 
We find that the absolute value of the energy shift
in the $J/\psi$-$\pi$ channel of Fig.~\ref{dEVDep}
decreases monotonically from $L=24$ to 48,
which is consistent with the leading 
$1/L^3$ behavior shown in Eq.~(\ref{largeL}). 
On the other hand,  the energy shift in the $J/\psi$-$\rho$ and $J/\psi$-$N$ 
channels do not show such monotonic decrease due to 
rather small values of $\Delta E$ at $L=24$.
This implies that the lattice size $L=24$ ($La \sim 1.6$ fm)
is too small to satisfy the condition $L> 2R \sim 2 a_0$ so that the
 convergence of the power series in Eq.~(\ref{largeL}) is questionable
  or even the phase-shift formula in Eq.~(\ref{phaseAC}) itself is invalid.
 Thus, in the $J/\psi$-$\rho$ and $J/\psi$-$N$ channels, 
 we do not take the results of $L=24$ as physical in the 
 following analyses.

\subsection{Volume dependence of scattering lengths}
\label{subsec:vol-dep-a}
 
\begin{figure}[htbp]
\begin{center}
\includegraphics[scale=0.33]{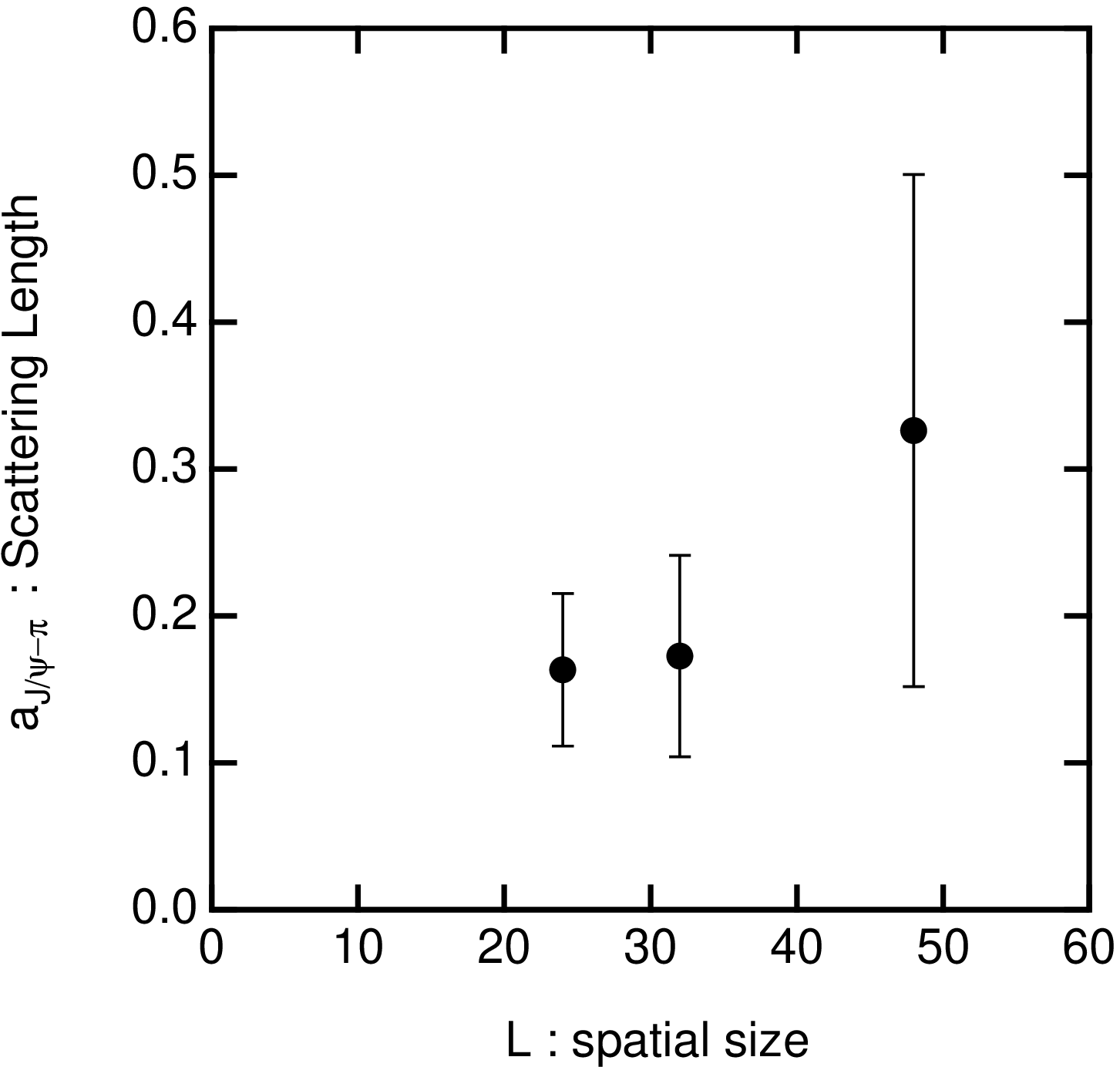}
\includegraphics[scale=0.33]{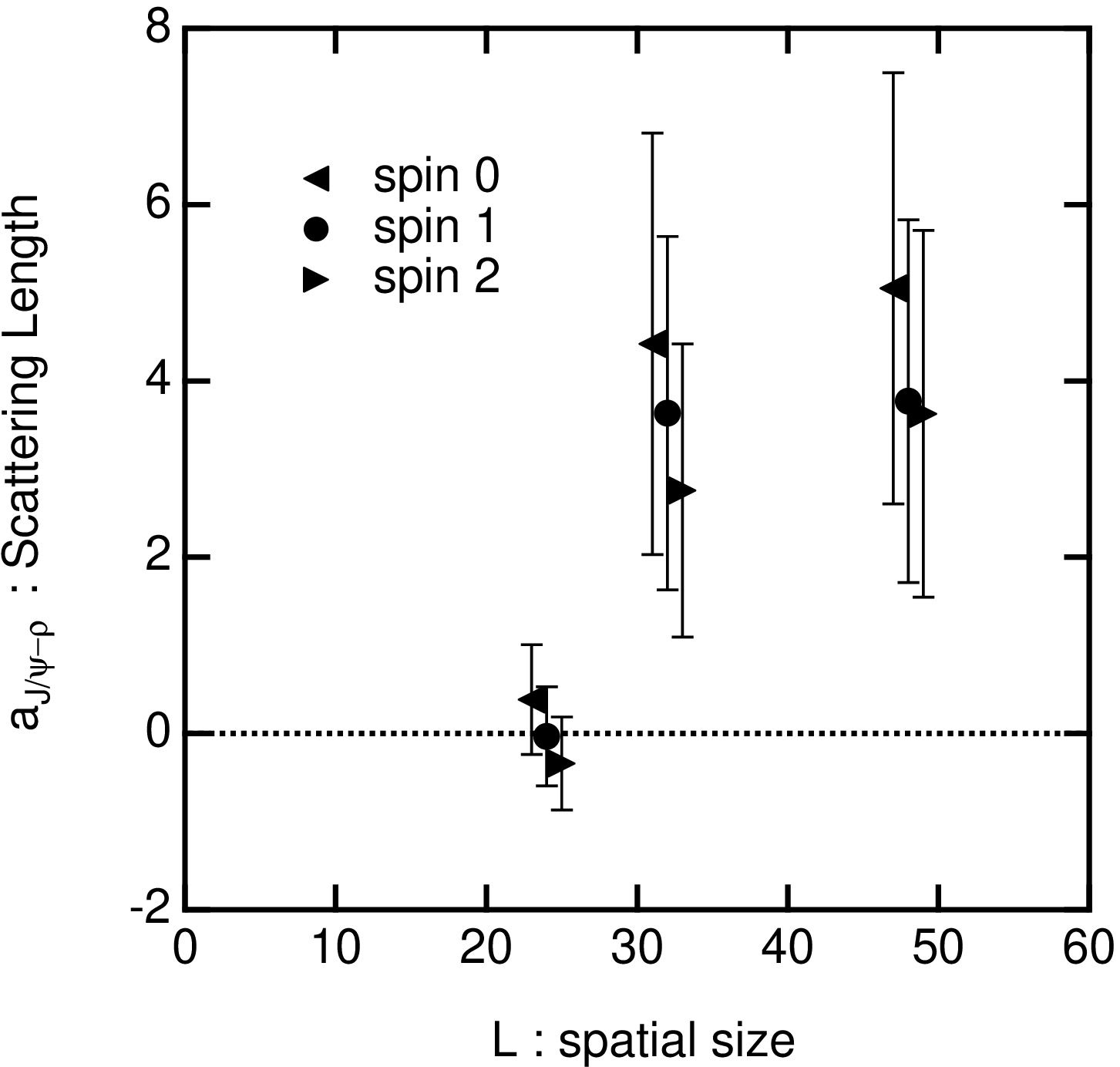}
\includegraphics[scale=0.33]{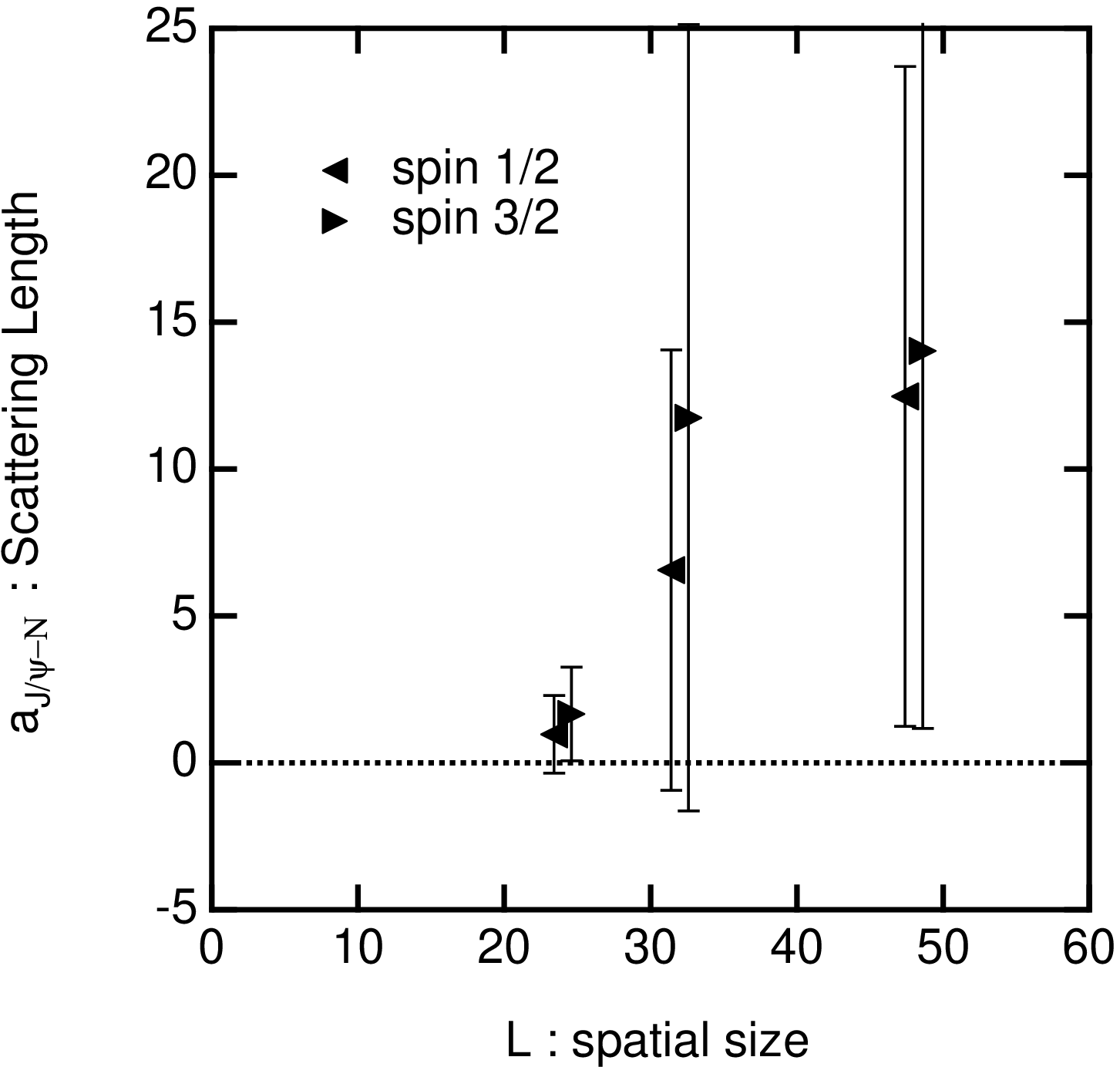}
\caption{
The scattering lengths as a function of the spatial size $L$ in lattice units
 for physical pion mass ($M_{\pi}=140$ MeV).
Left (middle, right) panel is for the $J/\psi$-$\pi$ ($J/\psi$-$\rho$, 
$J/\psi$-$N$) channel. 
}
\label{SLVDep}
\end{center}
\end{figure}

In the left panel of Fig.~\ref{SLVDep}, we show the volume dependence of
the $s$-wave scattering length $a_0^{J/\psi\textit{\rm -}\pi}$
at the physical point ($M_\pi=$140 MeV)
 obtained by applying the phase-shift formula (\ref{phaseAC}) to $\Delta E$ 
 in Fig.~\ref{dEVDep}.
 The horizontal axis is the spatial size $L$ and the vertical axis is the 
 scattering length.
 By taking into account the error bars for $L=48$ 
 due to the limited statistics, we do not find appreciable $L$
  dependence for the $J/\psi$-$\pi$ scattering lengths.
 Also, the scattering length is positive, which implies attraction 
between $J/\psi$ and $\pi$ at low energy.

In the middle panel of Fig.~\ref{SLVDep}, we show the volume dependence of 
the $s$-wave scattering length $a_0^{J/\psi\textit{\rm -}\rho}$
 obtained by Eq.~(\ref{phaseAC}).
 Within the error bars, different spin states have similar scattering lengths.
 The statistics in the $L=24$ case is 
  much better than $L=32$ and 48.  
  However, the $L=24$ data is contaminated
   by significant finite size effect
   as discussed in Sec.~\ref{sec:volume-dE} on the basis of Fig.~\ref{dEVDep}. 
 The scattering lengths for $L=32$ and 48 show
  the positive values.

In the right panel of Fig.~\ref{SLVDep}, we show the volume dependence of 
the $J/\psi$-$N$ scattering length $a_0^{J/\psi\textit{\rm -}N}$
obtained by Eq.~(\ref{phaseAC}).
Similar to the case in the $J/\psi$-$\rho$ channel,
we do not find appreciable spin dependence within the error bars.
 Here the same discussion as the $J/\psi$-$\rho$ channel applies 
 to the $L=24$ data.
 Although we need to increase statistics to make definite conclusion,
 the scattering length in $J/\psi$-$N$ channel for $L=32$ and 48
 is positive and is considerably larger than that in 
the $J/\psi$-$\pi$ channel.

In Table~\ref{Table_SL},
 we summarize the $s$-wave scattering lengths in lattice units
 at the physical point ($M_\pi$=140 MeV) and at the chiral limit ($M_{\pi}=0$).
 The table shows a clear tendency,
 \begin{eqnarray}
 0 < a_0^{J/\psi\textit{\rm -}\pi}   \ll 
 a_0^{J/\psi\textit{\rm -}\rho}  <
 a_0^{J/\psi\textit{\rm -}N}  .
 \end{eqnarray}
 This indicates that (i) there is always attraction between
$J/\psi$ and light hadrons, and (ii) there may be
some relation between the strength of the attraction and the 
 light-hadron properties such as the spatial size of light hadrons,
  or the number of constituent quarks in the light hadrons.
 To clarify the point (ii) further, it will be
  necessary not only to make high statistics simulations but also
 to carry out systematic calculations for the interactions of $J/\psi$ with various light hadrons. 
 It is also worth mentioning here that 
 the $s$-wave $J/\psi$-$\pi$ scattering length obtained at each volume 
is consistent with zero in the chiral limit within 1 or 2 standard deviations.
 This is in accordance with the current algebra result discussed in Eq.~(\ref{ChPT}).

\subsection{Scattering lengths in the physical units}   

In Table~\ref{Table_SLphys}, the results of the $s$-wave scattering lengths, 
$a_0^{J/\psi\textit{\rm -}h}$, for each $h$ with different spin states
are tabulated, where ``SAV" implies the results after the spin average.
We show the results of two different
methods to extract the scattering length from the energy shift.

 ``PSF (=Phase Shift Formula)" implies the method by using the 
 phase-shift formula in Eq.~(\ref{phaseAC}) without
 making the $1/L$-expansion. This is what we utilized to
 extract the numbers in Table~\ref{Table_SL}.
 In this case, the final results of the scattering lengths
 are obtained by an average over the scattering lengths
 obtained in different volumes.
  In an alternative method indicated by ``LLE (=Leading large-$L$ Expansion)"
  we take the leading $1/L$ formula in  Eq.~(\ref{largeL}),
 $\Delta E(L) = A \cdot L^{-3}$, 
 and calculate $A$ by
  fitting the volume dependence of the energy shift.
 The scattering length is then extracted as $a_0^{J/\psi\textit{\rm -}h} = -  \mu A/2\pi$. 
The values of $A$ obtained through this method are summarized
in Table~\ref{Table_Ampl}.
 Theoretically,  PSF is more rigorous than LLE.
Nevertheless, the latter is 
 useful to estimate a systematic error caused 
by a finite $L$. 

As we have discussed in Sections \ref{sec:volume-dE} and \ref{subsec:vol-dep-a},
we know that $L=24$ is not large enough to extract
any useful information in the 
$J/\psi$-$\rho$ and $J/\psi$-$N$ channels.
Therefore, in both PSF and LLE, we use only the data for $L=32$ and 48 in
the $J/\psi$-$\rho$ and $J/\psi$-$N$ channels, while
we use all data ($L=24, 32 $ and 48) in the $J/\psi$-$\pi$ channel. 

As seen in Table~\ref{Table_SLphys},
the results obtained from two analysis (PSF and LLE) agree with
each other within the errors. Especially, in the $J/\psi$-$\pi$ case,
both estimates are fairly consistent with each other.
The table also shows the elastic cross sections at the
threshold given by $\sigma_{\rm el} = 4\pi a_0^2$.
 We find that the $s$-wave scattering length in the $J/\psi$-$\pi$ 
channel is quite small, 0.0119(39) fm, but definitely positive.
Therefore, we conclude that the $J/\psi$-$\pi$ interaction is 
attractive at least at low energy.
The soft pion theorem, where implies that
the pion should decouple from any other hadrons in the chiral limit 
($M_{\pi}=0$), is an account for 
such smallness of the $s$-wave scattering length.
The $J/\psi$-$\pi$ elastic cross section at the threshold
is evaluated as $\sigma_{\rm el}^{J/\psi\textit{\rm -}\pi}
=0.018^{+0.013}_{-0.010}$ mb.
 This is consistent with a phenomenological analysis 
based on short distant QCD by Fujii and Kharzeev~\cite{FK}. 
  
Although statistical errors in $J/\psi$-$\rho$ and $J/\psi$-$N$ channels
are quite large, we find that $s$-wave scattering lengths in these 
channels are likely to be positive
and are at least an order of magnitude larger than the pion case.
As for the $J/\psi$-$N$ channel, 
 the scattering length from QCD sum rules by Hayashigaki \cite{AH},
gluonic van der Waals interaction by Brodsky {\it et al.}~\cite{BG97},
and  the QCD multipole expansion by Voloshin {\it et al.}~\cite{SV}
show that the $s$-wave scattering length is about
  0.1 fm, 
 0.25 fm, 
 and 0.37 fm or larger, respectively. 
 Our central value of the scattering length 0.71 fm from PSF (0.39 fm from LLE)
 is comparable or even larger than those estimates, but
we need to increase statistics of our data to draw
solid comparison.

\subsection{Possible contaminations from channel mixings}

Before closing the discussion on $J/\psi$-hadron scattering lengths,
we comment on possible contaminations
 from the $\eta_c$-$h$ and $D{\bar D}$ states.
 If there were open channels with lower energy than the 
  $J/\psi$-hadron system at the threshold,
  the L\"uscher's formula for extracting the scattering 
  phase shift from $\Delta E$ in the desired channel is not applicable.  
  
The $J/\psi$-$\pi$ system is free from the
 contamination of the $\eta_c$-$\pi$ subthreshold state.
 This is because the $s$-wave $\eta_c$-$\pi$ state has different
  total spin from the $s$-wave $J/\psi$-$\pi$  state.
 On the other hand, for the $J/\psi$-$\rho$ case, 
we cannot exclude the possible contamination  
of the $\eta_c$-$\rho$ state to the spin-1 $J/\psi$-$\rho$ state
and that of the $D{\bar D}$ state to the spin-0 $J/\psi$-$\rho$ state.
Only the highest spin state (spin-2) is definitely free from such contaminations. 
This is true also for the case of $J/\psi$-$N$ system:
The highest spin state (spin-3/2) is free from the 
 contamination of the $\eta_c$-$N$ state.

Thus, strictly speaking, the spin-1 $J/\psi$-$\pi$, the spin-2 $J/\psi$-$\rho$ and
the spin-3/2 $J/\psi$-$N$ states are the safe channels in determining
the $s$-wave scattering phase shifts from the L\"uscher's formula,
although in our simulations we do not find any appreciable difference 
among different spin channels
within the error bars.

\subsection{$\eta_c$-hadron interactions}
\label{sec:etac} 

The scattering length of $\eta_c$ with light hadrons can
 be performed in exactly the same way as the $J/\psi$ case by
 using the conventional $\eta_c$ interpolating operator as in Sec.~\ref{def-IntOp}:
\begin{eqnarray}
{\cal O}^{\eta_c}({\vec x},t)=\bar{c}_a({\vec x},t) \gamma_{5} c_a({\vec x},t).
\end{eqnarray}
In the present paper,
 we neglect the disconnected diagrams in both
 the two-point function of $\eta_c$ and the related four-point functions
 as we have done in the case of $J/\psi$.
 From the theoretical point of view,
 neglecting  the disconnected diagrams in the 
 $\eta_c$ channel is less justified than that in the $J/\psi$ channel.
 On the other hand,
 it is numerically observed that the contributions from the disconnected
diagrams are  negligibly small for the charm quark even in the 
pseudoscalar channel~\cite{UK,TARO}.
 
In our simulations the mass difference between the $J/\psi$ and $\eta_c$ is 
about 30 MeV,
 which is considerably smaller than the experimental value, 116 MeV. 
 It is known that the hyperfine splitting is sensitive to 
 the quenched approximation and the leading discretization error of 
 the Wilson fermion action ~\cite{UK,TARO}. 
 Therefore, the spin-dependent observables in our simulations
 should be taken with caution.
 
\begin{figure}[htbp]
\begin{center}
\includegraphics[scale=0.33]{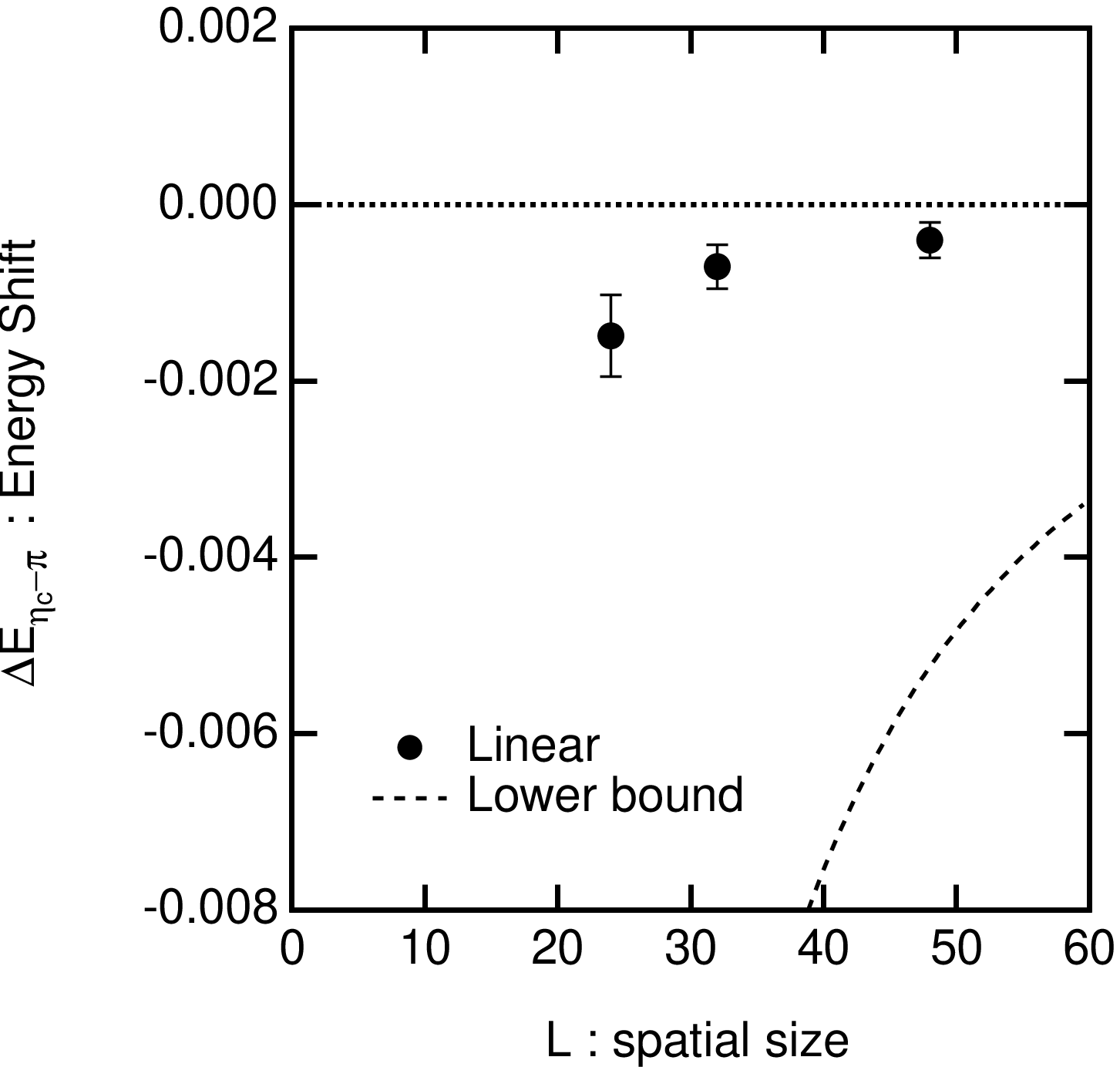}
\includegraphics[scale=0.33]{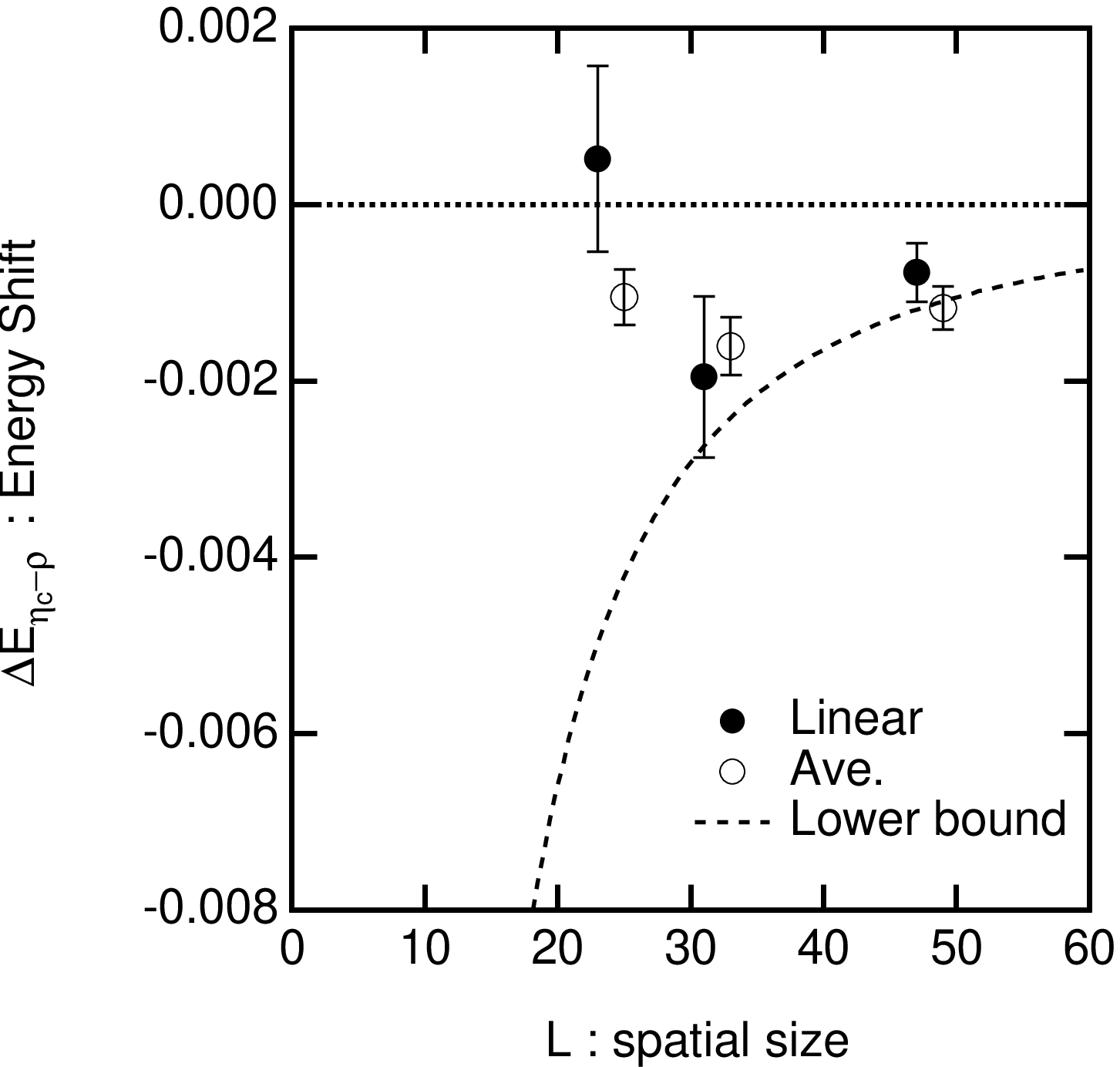}
\includegraphics[scale=0.33]{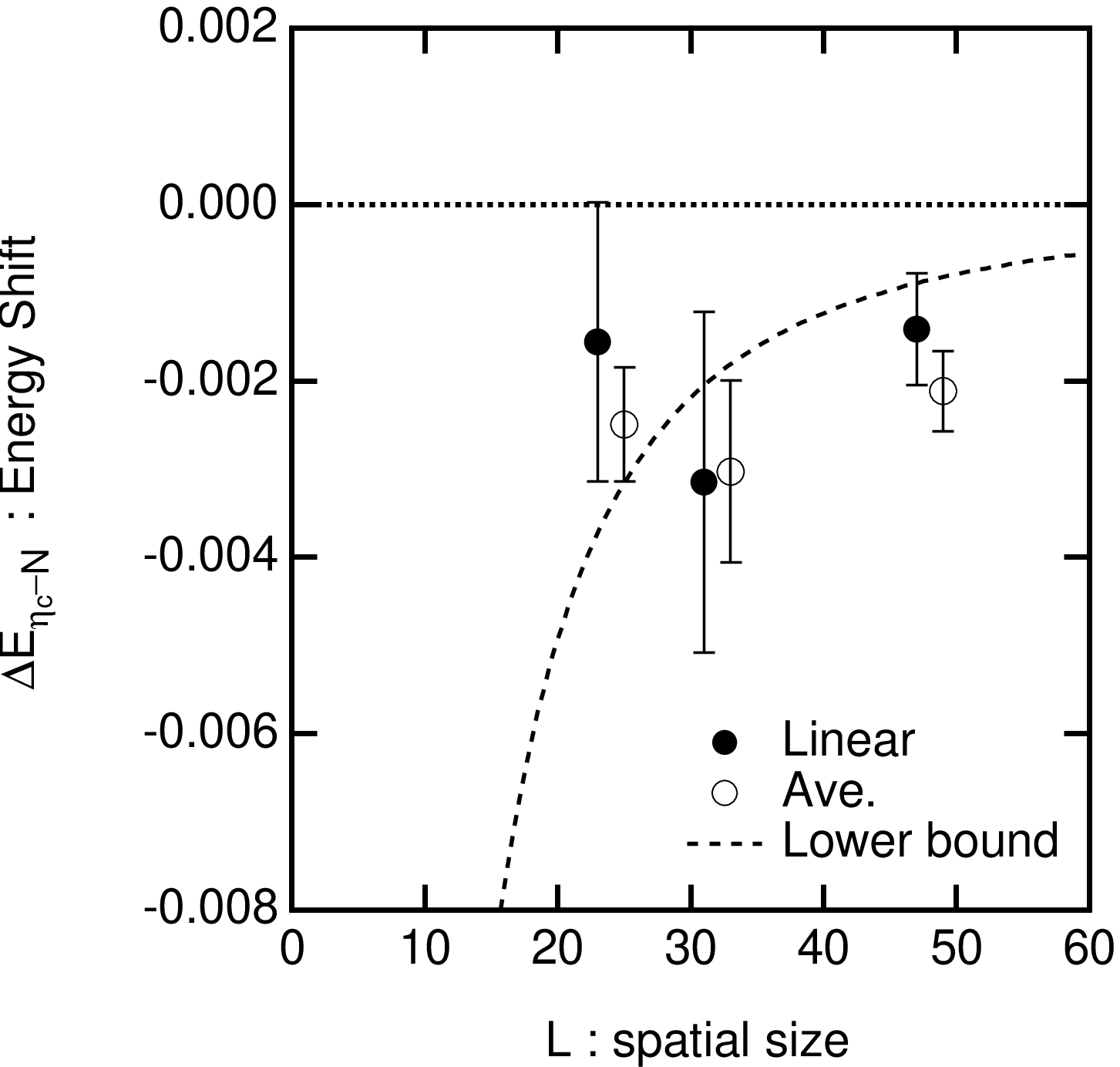}
\caption{
The energy shifts $\Delta E$ as a function of spatial size $L$ in lattice units.
Left (middle, right) panel is for the $\eta_c$-$\pi$ ($\eta_c$-$\rho$, 
$\eta_c$-$N$) channel. 
Full and open circles represent the values obtained from the linear chiral-extrapolation
to the physical point, and the weighted average, respectively. 
The dashed curves show the lower boundary for the convergence of the large-$L$ expansion
of $\Delta E$.
}\label{dEVDepEtac}
\end{center}
\end{figure}

\begin{figure}[htbp]
\begin{center}
\includegraphics[scale=0.33]{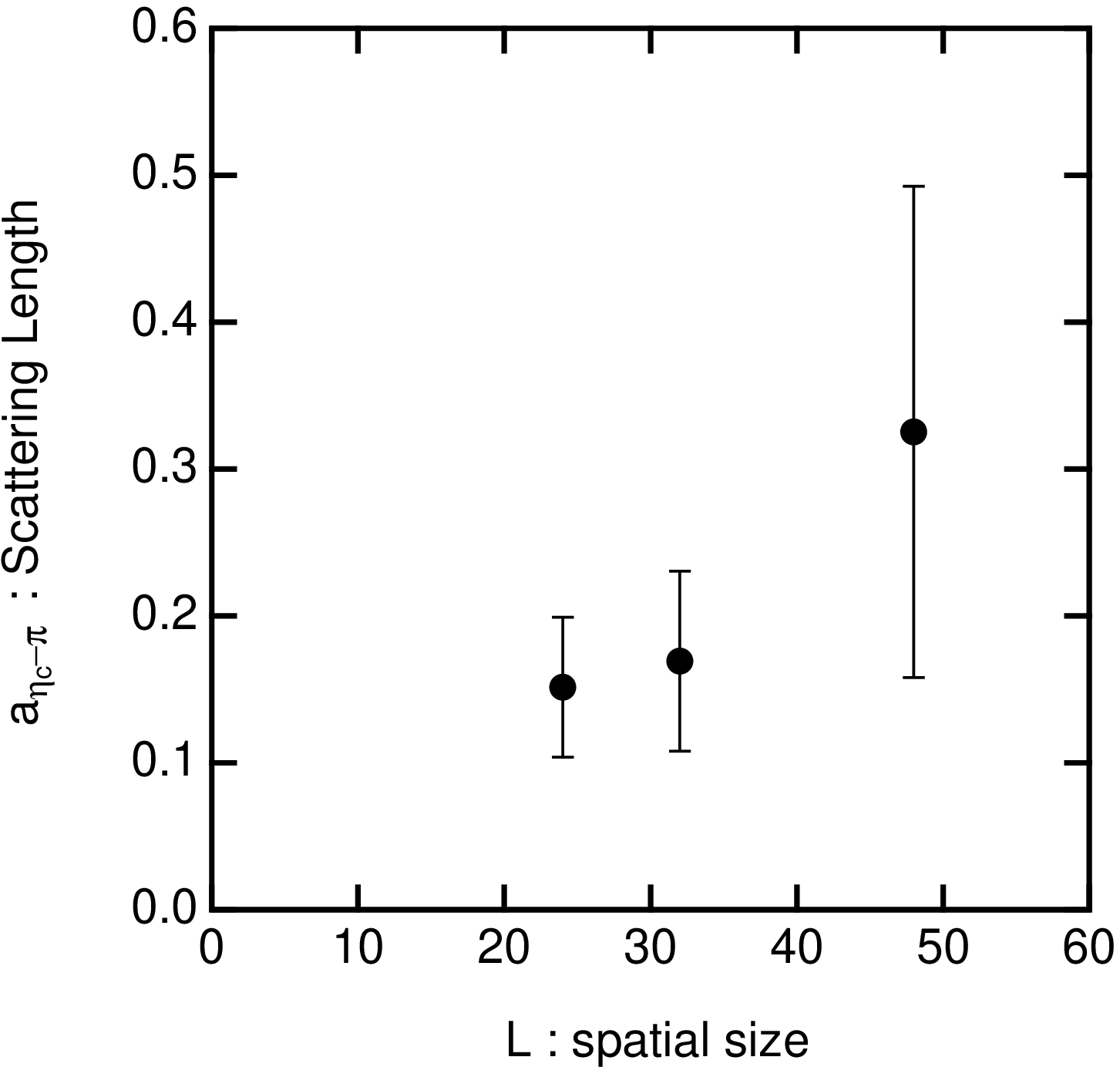}
\includegraphics[scale=0.33]{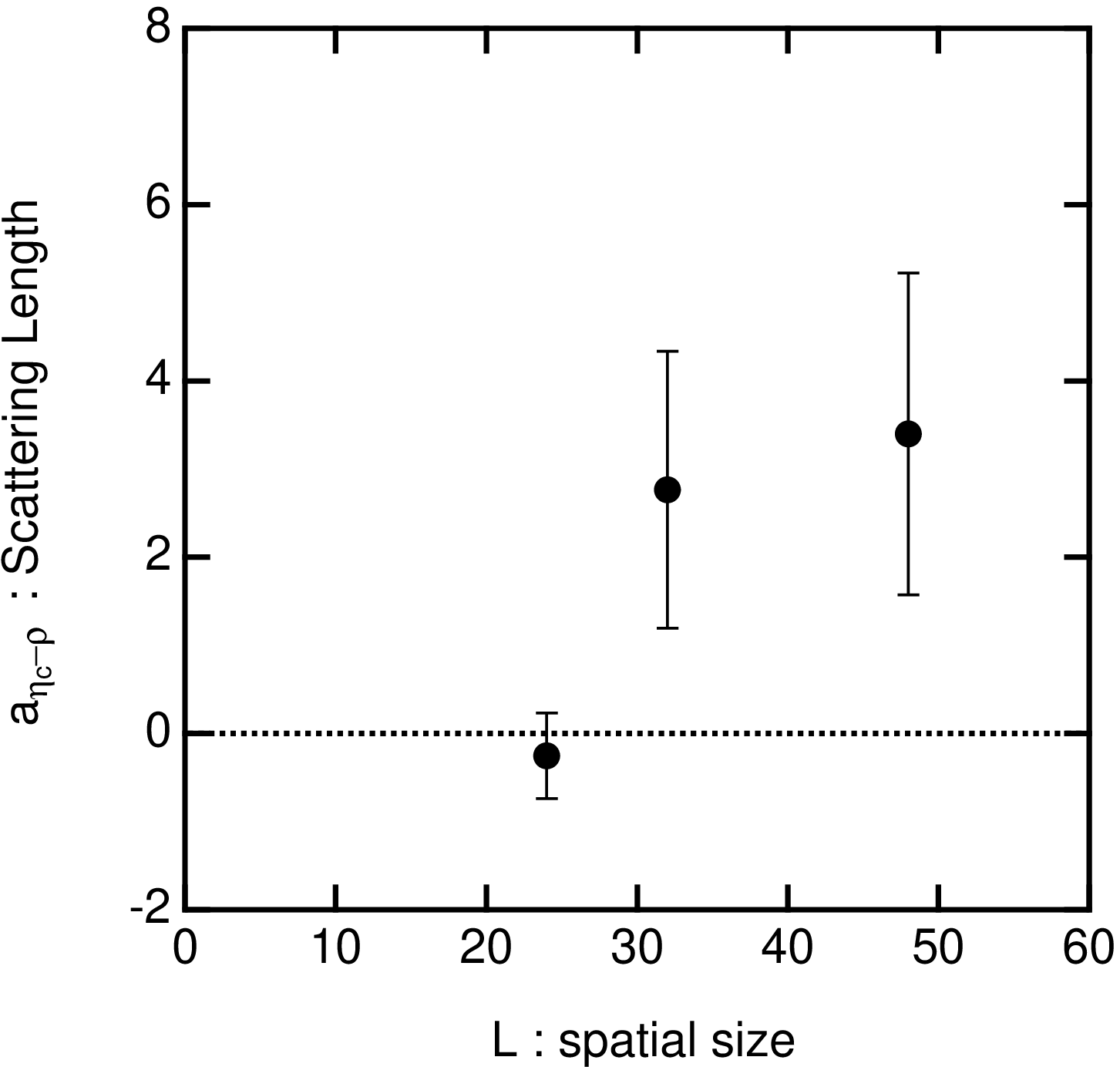}
\includegraphics[scale=0.33]{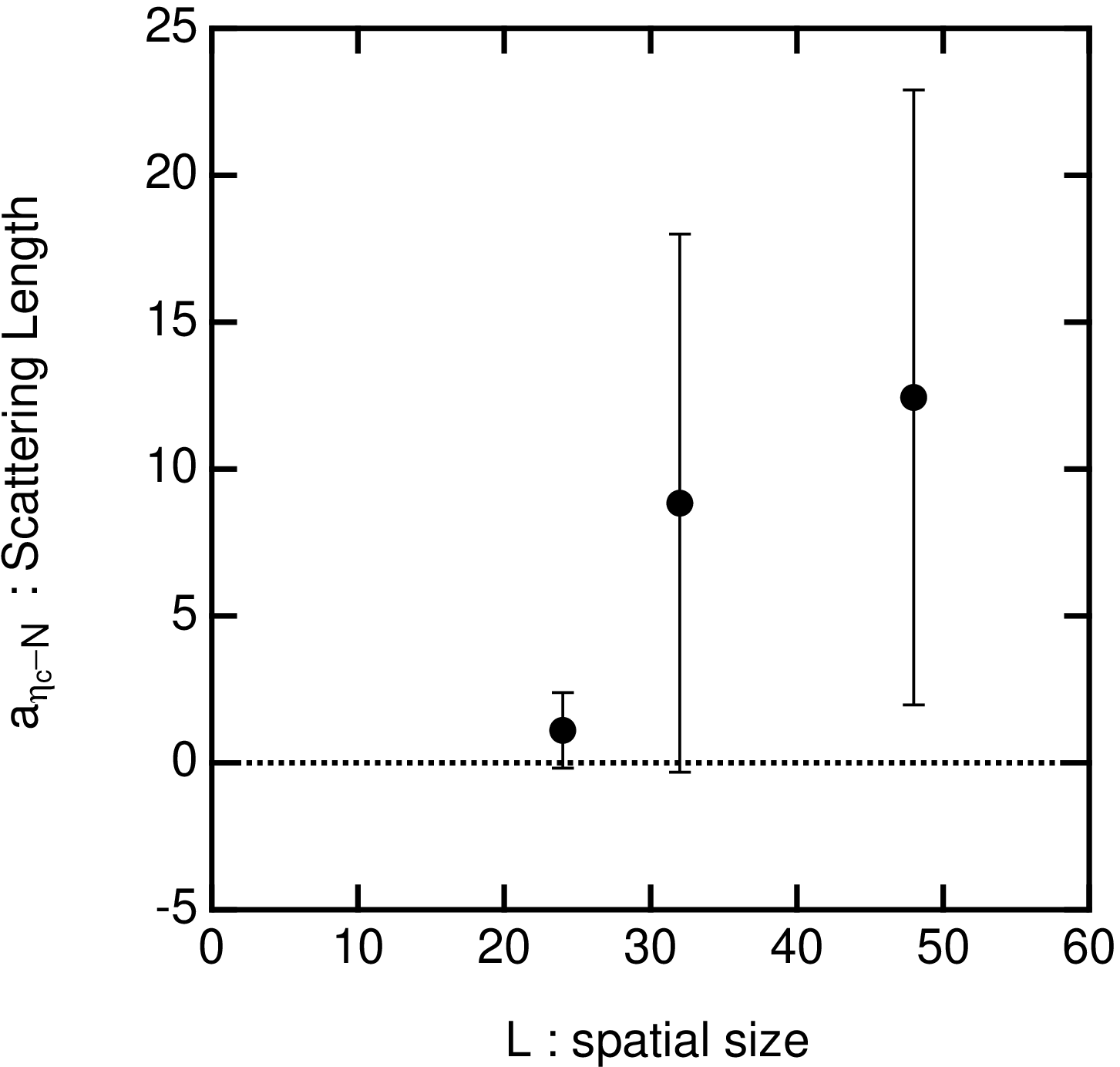}
\caption{
The scattering lengths as a function of the spatial size $L$ in lattice units
 for the physical pion mass ($M_{\pi}=140$ MeV).
Left (middle, right) panel is for the $\eta_c$-$\pi$ ($\eta_c$-$\rho$, 
$\eta_c$-$N$) channel. 
}
\label{SLVDepEtac}
\end{center}
\end{figure}

 Under these reservations,
 the volume dependences of the energy shift and the scattering length
 are shown in Figs.~\ref{dEVDepEtac} and \ref{SLVDepEtac}, respectively.
 Qualitative features of the figures are similar to  those
  in the  $J/\psi$ case: The energy shift is definitely negative
in all three channels at $L= 32$ and 48. 
Also, we find $0 < a_0^{\eta_c\textit{\rm -}\pi} \ll a_0^{\eta_c\textit{\rm -}\rho}
< a_0^{\eta_c\textit{\rm -}N}$.
In the $\eta_c$-$\rho$ and $\eta_c$-$N$ channels, considerable finite $L$ effect 
may exist at $L=24$. All analyses are made in the same way 
as the case of $J/\psi$. 
Table~\ref{Table_SL_etac} contains
two types of the chirally extrapolated value of the $s$-wave scattering length
in all channels for each volume. In Table~\ref{Table_SLphys_etac}, we
compare results obtained from two different analyses (PSF and LLE). 
 The coefficients $A$ in LLE are tabulated in Table~\ref{Table_Ampl_etac}.

 The $s$-wave $\eta_c$-$\pi$ scattering length turns out to be
0.0113$\pm$0.0035 fm and the corresponding 
 elastic cross section at the threshold is
  $\sigma^{\eta_c\textit{\rm -}\pi}_{\rm el}=0.016^{+0.011}_{-0.008}$ mb, 
both of which are very close to the values in the $J/\psi$-$\pi$ channel.
Although we have large error bars, the scattering lengths
 in the $\eta_c$-$\rho$ and $\eta_c$-$N$ channels are also comparable with those
  in the  $J/\psi$-$\rho$ and  $J/\psi$-$N$ channels.

\section{Summary and concluding remarks}
 
In this paper,
we have studied the interactions of 
$J/\psi$ and $\eta_c$ with light-hadrons at low energy
in quenched lattice QCD simulations.
We calculated the scattering lengths of
$J/\psi$-$\pi,\rho, N$ and  $\eta_c$-$\pi,\rho,N$ 
from the energy shifts $\Delta E$ of two hadrons
in a finite periodic box $L^3$ 
 by using the L\"uscher's phase-shift formula.
 We employed the lattice spacing $a \simeq (2.9\ {\rm GeV})^{-1}$,
 three light-quark masses $M_{\pi}/M_{\rho}\simeq 0.68, 0.83, 0.90$,
  and three lattice sizes $La\approx 1.6, 2.2$ and 3.2 fm. 
 Their values were utilized to perform analyses
 on quark-mass and lattice-size dependences
 of the energy shifts and the scattering lengths.
 
 We did not find appreciable quark-mass dependence of $\Delta E$
 in all channels, while in the $J/\psi$-$\pi$ channel
 it turned out that there is a tendency that the energy shift approaches
 zero in simultaneous chiral and large-volume limits 
 as indicated in the current algebra analysis.

As for the volume dependence of the energy shifts,
the $1/L^3$ behavior, which is expected from the asymptotic
expansion of the L\"uscher's formula in terms of $1/L$,
was seen in the $J/\psi$-$\pi$ and $\eta_c$-$\pi$ channels.
On the other hand, 
in the $J/\psi$-$\rho, N$ and $\eta_c$-$\rho, N$ channels,
the data at $L=24$ do not follow the 1/$L^3$ behavior~\cite{footnote1}.
Indeed, $La = 1.6$ fm is not large enough to
accommodate even a single hadron such as $N$~\cite{footnote2}. 

On the basis of the above observation, we adopted
the data for $L=24$, 32 and 48
in the $J/\psi(\eta_c)$-$\pi$ channel and the data for $L=32$ and 48
in the $J/\psi(\eta_c)$-$\rho$ and $J/\psi(\eta_c)$-$N$ channels.
Then we applied the L\"uscher's formula without $1/L$-expansion (which
 we call PSF) and with $1/L$-expansion (which we call LLE).
The resultant $s$-wave scattering lengths obtained from $\Delta E$
 show that (i) the interaction of charmonia with light hadrons
  is always attractive, (ii) the scattering lengths in the 
  $J/\psi(\eta_c)$-$\pi$ channels are quite small compared to other channels
  as consistent with the soft-pion theorem, and (iii) the scattering lengths in the
 $J/\psi(\eta_c)$-$\rho$ and $J/\psi(\eta_c)$-$N$ channels
 are at least an order of magnitude larger than those in the
 $J/\psi(\eta_c)$-$\pi$ channels.

The $s$-wave $J/\psi$-$\pi$ ($\eta_c$-$\pi$) scattering length is determined
as 0.0119$\pm$0.0039 fm (0.0113$\pm$0.0035 fm) and the corresponding
elastic cross section at the threshold
becomes $0.018^{+0.013}_{-0.010}$ mb ($0.016^{+0.011}_{-0.008}$ mb).
 On the other hand, the $J/\psi$-$N$ ($\eta_c$-$N$) spin-averaged scattering length  
is  0.71$\pm$0.48 fm (0.70$\pm$0.66 fm) which has still a large
statistical errors. Nevertheless,  our result in the $J/\psi$-$N$ channel 
may be compared with estimates from QCD sum rules ($\sim$ 0.1 fm),
 from the gluonic van der Waals interaction ($\sim$ 0.25 fm)
 and from the QCD multipole expansion ($\sim$ 0.37 fm or larger).  
        
 To have more quantitative understanding of the scattering lengths,
 in particular, of their magnitudes and spin dependences,
  we need to accumulate more statistics.
  Since our data for the $J/\psi(\eta_c)$-$\rho(N)$ channels indicate that
  the scattering length is rather large so that
  the large-$L$ expansion is not fully justified even for
  $L=32$ and 48. Therefore, it is important to go to
  larger volume to reduce the systematic errors due to the finite volume.     
  Simulations with dynamical quarks are also an important future direction
 to be explored for full knowledge of the charmonium interactions.

 
\subsection*{Acknowledgement}
 
We acknowledge to K.~Sasaki for beneficial discussions
and helping us to develop codes of the gauge fixing.
We thank for A.~Nakamura and his collaborators for giving us a chance to use 
their open source codes (Lattice Tool Kit \cite{LTK}).
S.S. thanks D.~Kharzeev and T.~Yamazaki for fruitful discussions.
This work was supported by the Supercomputer Projects No.110 (FY2004)
and No.125 (FY2005) of High Energy Accelerator Research
Organization (KEK). T.H. was supported in 
part by the Grants-in-Aid of the Japanese Ministry
of Education, Culture, Sports, Science, and Technology (No.~15540254).


\section*{Appendix A: Spin projection}
 
In the cases of the $s$-wave $J/\psi$-$\rho$ and $J/\psi$-$N$ scatterings, 
there are different spin states: Spin-0, 1 and 2 states for the $J/\psi$-$\rho$ and
spin-1/2 and 3/2 states for the $J/\psi$-$N$. Therefore, the appropriate spin
projections are required to disentangle each spin contribution from 
four-point correlation functions.

For the $J/\psi$-$\rho$ system, the four point function, Eq.~(\ref{Eq:FourPointRho}),
can be expressed by the orthogonal sum of spin-0, spin-1 and spin-2 components:
\begin{eqnarray}
G^{J/\psi \textit{{\rm -}} \rho}_{ij;kl}(t) 
= G^{0}(t)\hat{P}^{0}_{ij;kl} +G^{1}(t)\hat{P}^{1}_{ij;kl} +G^{2}(t)\hat{P}^{2}_{ij;kl}
\end{eqnarray}
with spin projection operators defined by
\begin{eqnarray}
  \hat{P}^{0}_{ij;kl} &=& \frac{1}{3}\delta_{ij}\delta_{kl},  \\
  \hat{P}^{1}_{ij;kl} &=& \frac{1}{2}(\delta_{ik}\delta_{jl}-\delta_{il}\delta_{jk}),  \\
  \hat{P}^{2}_{ij;kl} &=& \frac{1}{2}(\delta_{ik}\delta_{jl}+\delta_{il}\delta_{jk})
-\frac{1}{3}\delta_{ij}\delta_{kl}.
\end{eqnarray}
Respective spin-projected correlators are given as
%
%
\begin{eqnarray}
G^{0}(t) &=&  \frac{1}{3}\sum_{i,j=1}^3   G^{J/\psi \textit{{\rm -}} \rho}_{ii;jj}(t), \\
G^{1}(t) &=&  \frac{1}{6}\sum_{i,j=1}^3   \left( 
G^{J/\psi \textit{{\rm -}} \rho}_{ij;ij}(t)   -G^{J/\psi \textit{{\rm -}} \rho}_{ij;ji}(t) 
\right), \\
G^{2}(t) &=&  \frac{1}{10}\sum_{i,j=1}^3   \left(
G^{J/\psi \textit{{\rm -}} \rho}_{ij;ij}(t)    +G^{J/\psi \textit{{\rm -}} \rho}_{ij;ji}(t) 
-\frac{2}{3}G^{J/\psi \textit{{\rm -}} \rho}_{ii;jj}(t)
\right),
\end{eqnarray}
where indices $i$ and $j$ should be summed over all spatial directions.

The four point function for the $J/\psi$-$N$ system, Eq.~(\ref{Eq:FourPointNuc}),
can be also decomposed into spin-1/2 and spin-3/2 components as
%
%
\begin{eqnarray}
G^{J/\psi \textit{{\rm -}} N}_{ij}(t) 
= G^{1/2}(t)\hat{P}^{1/2}_{ij} +G^{3/2}(t)\hat{P}^{3/2}_{ij}.
\end{eqnarray}
Here, spin projection operators for spin-1/2 and spin-3/2 are given by
%
%
\begin{eqnarray}
  \hat{P}^{1/2}_{ij} &=&  \frac{1}{3}\gamma_i \gamma_j, \\
  \hat{P}^{3/2}_{ij} &=&  \delta_{ij} -\frac{1}{3}\gamma_i \gamma_j.
\end{eqnarray}
Then, each spin part can be projected out as
%
%
\begin{eqnarray}
G^{1/2}(t) &=&    \sum_{i,j=1}^3    \hat{P}^{1/2}_{ij}   G^{J/\psi \textit{{\rm -}} N}_{ji}(t)
=\frac{1}{3}\sum_{i,j=1}^3 \gamma_{i}\gamma_{j}  G^{J/\psi \textit{{\rm -}} N}_{ji}(t),
\\
G^{3/2}(t) &=&    \frac{1}{2}\sum_{i,j=1}^3    \hat{P}^{3/2}_{ij}   G^{J/\psi \textit{{\rm -}} N}_{ji}(t)
= \frac{1}{2}\sum_{i=1}^{3}G^{J/\psi \textit{{\rm -}} N}_{ii}(t)
-\frac{1}{6}\sum_{i,j=1}^{3}\gamma_{i}\gamma_{j} G^{J/\psi \textit{{\rm -}} N}_{ji}(t),
\end{eqnarray}
where indices $i$ and $j$ are also summed over all spatial directions. 
Recall that respective contributions $G^{1/2}(t)$ and $G^{3/2}(t)$ 
possess non-trivial Dirac structure. To extract the particle 
 contribution, we need to take a trace with the projection operator $(1+\gamma_4)/2$~\cite{SBO}.


\newpage

%
%
\begin{table}[htdp]
\caption{Simulation parameters in this study.
The Sommer parameter $r_0=0.5$ fm is used 
to fix the scale~\cite{Sscale1,Sscale2}.}
\begin{ruledtabular}
\begin{tabular}{cccccccc}
$\beta$ &$a$ [fm]&$a^{-1}$ [GeV]&&Lattice size ($L^3\times T$)  
& $\sim La$ [fm]&& Statistics \\\hline
6.2         &$0.06775$&2.913&&$24^3\times48 $&1.6          && 161\\
               &               &&&$32^3\times 48$&2.2          &&169\\
               &               &&&$48^3\times 48$&3.2          &&53 \\
\end{tabular}
\end{ruledtabular}
\label{lConst}
\end{table}
%

%
%
\begin{table}[htdp]
\caption{
Fitted masses of pseudoscalar, vector and nucleon states in lattice units.
We perform a covariant single exponential fit to two-point functions
of each hadron in respective fitting ranges.
The vector-meson mass at $\kappa=$0.1360 in the physical unit 
is close to the mass of $J/\psi(3097)$ and hence the value is reserved
for a charm quark. Other three hopping parameters are for
light quarks. The box-to-point quark propagators 
 are used in the present study, while the 
 point-to-point quark propagators are used in
 Ref.~\cite{KSSS} with different gauge configurations.
 The hadron masses in two approaches agree well with each other.
}\label{table-mass}
\begin{ruledtabular}
\begin{tabular}{cc|cccc}
$L^3\times T$& Fitting range &
$\kappa$ &
$aM_\pi$&$aM_\rho$&$aM_N$ 
\\\hline
$24^3\times 48$& [22,31]
&0.1360 &  1.017(1)& 1.027(1)& 1.584(2)  \\
&&0.1489 &  0.415(1)& 0.461(2)& 0.722(4)  \\
&&0.1506 &  0.314(2)& 0.383(3)& 0.589(5)  \\
&&0.1520 &  0.213(2)& 0.322(6)& 0.467(10)  \\\hline
$32^3\times 48$ & [25, 34]
&0.1360 &  1.020(1)& 1.030(1)& 1.594(2)  \\
&&0.1489 &  0.416(1)& 0.463(1)& 0.723(4)  \\
&&0.1506 &  0.315(1)& 0.383(2)& 0.592(4)  \\
&&0.1520 &  0.213(1)& 0.317(4)& 0.473(8)  \\\hline
$48^3\times 48$ & [17,26]
&0.1360 &  1.016(1)& 1.026(1)& 1.581(3)  \\
&&0.1489 &  0.415(1)& 0.460(2)& 0.719(4)  \\
&&0.1506 &  0.314(1)& 0.380(2)& 0.586(4)  \\
&&0.1520 &  0.212(1)& 0.311(3)& 0.465(5)  \\\hline
\end{tabular}
\end{ruledtabular}
\end{table}
%

%
%
\begin{table}[htdp]
\caption{Energy shifts $\Delta E$
 in all $J/\psi$-$h$ channels on the lattice with $L=24$. 
The energy shifts are obtained from the ratios $R_{J/\psi \textit{\rm -}h}$ 
by the single exponential fit. The fitting range is chosen to be $22\le t \le31$ 
for all channels so as to be the same as that employed for spectroscopy 
of single hadrons for $L=24$.
The labels ``phys." and ``chiral" show the  values extrapolated to the
physical point ($M_{\pi}=140$ MeV) and to the chiral limit, respectively,
 with the linear quark-mass dependence in Eq.~(\ref{chiext}).
}
\begin{ruledtabular}
\begin{tabular}{ccccccccccc}
$\kappa$ &$\quad$& 
$a\Delta E_{J/\psi  \textit{\rm -}\pi}$ &&
 $a\Delta E_{J/\psi \textit{\rm -}\rho}^{0}$ &
 $a\Delta E_{J/\psi \textit{\rm -}\rho}^{1}$ &
 $a\Delta E_{J/\psi \textit{\rm -}\rho}^{2}$ &&
  $a\Delta E_{J/\psi \textit{\rm-}N}^{1/2}$ &
  $a\Delta E_{J/\psi \textit{\rm-}N}^{3/2}$ &
 \\\hline
0.1489&& -0.0013(4)&&-0.0020(4)  &-0.0016(4)&-0.0012(4)&&     -0.0028(9)& -0.0032(9)\\
0.1506&& -0.0014(4)&&-0.0016(6)  &-0.0012(6)&-0.0007(6)&&     -0.0023(10)&-0.0028(11)\\
0.1520&& -0.0016(5)&&-0.0006(13)&0.0003(13)&0.00010(13)&& -0.0016(16)&-0.0023(17)\\\hline
phys.&&   -0.0016(5)&&-0.0008(12)&0.0000(12)&0.0007(12)&&   -0.0014(17)&-0.0022(17)\\
chiral&&   -0.0016(5)&&-0.0008(12)&0.0001(12)&0.0007(12)&&   -0.0014(17)&-0.0022(18)\\
\end{tabular}
\end{ruledtabular}
\label{Table:EsftL24}
\end{table}
%

%
%
\begin{table}[htdp]
\caption{Energy shifts for all $J/\psi$-$h$ channels on the lattice with $L=32$.
The fitting range is chosen to be $25\le t \le34$  for all channels.
The labels are the same as in Table~\ref{Table:EsftL24}.
}
\begin{ruledtabular}
\begin{tabular}{ccccccccccc}
$\kappa$ &$\quad$& 
$a\Delta E_{J/\psi \textit{\rm -}\pi}$ &&
$a\Delta E_{J/\psi \textit{\rm -}\rho}^{0}$ &
$a\Delta E_{J/\psi \textit{\rm -}\rho}^{1}$ &
$a\Delta E_{J/\psi \textit{\rm -}\rho}^{2}$ &&
$a\Delta E_{J/\psi \textit{\rm -}N}^{1/2}$ &
$a\Delta E_{J/\psi \textit{\rm -}N}^{3/2}$ &
 \\\hline
0.1489&&-0.0010(3) &&-0.0021(5)   &-0.0019(5)  &-0.0016(5)    &&-0.0032(15)&-0.0035(16)\\
0.1506&&-0.0009(3) &&-0.0023(6)   &-0.0021(6)  &-0.0017(6)    &&-0.0028(14)&-0.0033(16)\\
0.1520&&-0.0008(3) &&-0.0027(11)&-0.0023(11)&-0.0018(10) &&-0.0029(19)&-0.0039(22)\\\hline
phys.    &&-0.0007(3) &&-0.0028(11)&-0.0024(10)&-0.0019(10) &&-0.0027(19)&-0.0036(21)\\
chiral   &&-0.0007(3) &&-0.0028(11)&-0.0024(10)&-0.0020(10) &&-0.0027(19)&-0.0036(22)\\
\end{tabular}
\end{ruledtabular}
\label{Table:EsftL32}
\end{table}
%

%
%
\begin{table}[htdp]
\caption{Energy shifts for all $J/\psi$-$h$ channels on the lattice with $L=48$.
The fitting range is chosen to be $17\le t \le26$  for all channels.
The labels are the same as in Table~\ref{Table:EsftL24}.}
\begin{ruledtabular}
\begin{tabular}{ccccccccccc}
$\kappa$ &$\quad$& 
$a\Delta E_{J/\psi \textit{\rm -}\pi}$ &&
$a\Delta E_{J/\psi \textit{\rm -}\rho}^{0}$ &
$a\Delta E_{J/\psi \textit{\rm -}\rho}^{1}$ &
$a\Delta E_{J/\psi \textit{\rm -}\rho}^{2}$ &&
$a\Delta E_{J/\psi \textit{\rm -}N}^{1/2}$ &
$a\Delta E_{J/\psi \textit{\rm -}N}^{3/2}$ &
 \\\hline
0.1489&&-0.0010(3) &&-0.0015(4)&-0.0014(4)&-0.0013(4) &&-0.0025(8)&-0.0026(8)\\
0.1506&&-0.0008(2) &&-0.0014(3)&-0.0013(3)&-0.0012(3) &&-0.0020(6)&-0.0022(6)\\
0.1520&&-0.0005(2) &&-0.0011(3)&-0.0009(3)&-0.0009(3) &&-0.0017(6)&-0.0018(6)\\\hline
phys.    &&-0.0004(2) &&-0.0010(4)&-0.0008(4)&-0.0008(4) &&-0.0014(7)&-0.0015(7)\\
chiral   &&-0.0004(2) &&-0.0010(4)&-0.0008(4)&-0.0008(4) &&-0.0014(7)&-0.0015(7)\\
\end{tabular}
\end{ruledtabular}
\label{Table:EsftL48}
\end{table}%

%
%
\begin{table}[htdp]
\caption{Results of the chirally extrapolated values of 
$s$-wave scattering lengths in lattice units for all $J/\psi$-$h$ channels.
The labels are the same as in Table~\ref{Table:EsftL24}.
}
\begin{ruledtabular}
\begin{tabular}{cc|c|ccc|cc}
&&
$ a_0^{J/\psi \textit{\rm -}\pi}$ &
$ a_0^{J/\psi \textit{\rm -}\rho}$ &&&
$ a_0^{J/\psi \textit{\rm -}N}$ &
\\
$L^3\times T$ & &
 &
spin-0 &
spin-1 &
spin-2 &
spin-1/2 &
spin-3/2 
 \\\hline
$24^3\times 48$ &
phys.& 0.16(5)         &0.38(62)&-0.03(56)&-0.34(53)& 0.97(1.32)&1.66(1.59)\\
&chiral& -0.0029(18)&0.38(62)&-0.03(56)&-0.34(53)& 0.96(1.31)&1.65(1.57)\\\hline
$32^3\times 48$ &
phys.   & 0.17(7)        &4.4(2.4)&3.6(2.0)&2.8(1.7)& 6.6(7.5)&11.7(13.4)\\
&chiral& -0.0013(10)&4.4(2.4)&3.6(2.0)&2.7(1.6)& 6.5(7.4)&11.5(13.0)\\\hline
$48^3\times 48$ &
phys.   & 0.33(17)     &5.1(2.4)&3.8(2.1)&3.6(2.1)& 12.5(11.2)&14.0(12.8)\\
&chiral& -0.0014(15)&5.0(2.4)&3.7(2.0)&3.6(2.1)&  12.3(11.0)&13.8(12.5)
\end{tabular}
\end{ruledtabular}
\label{Table_SL}
\end{table}
%


%
%
\begin{table}[t]
\caption{
The $s$-wave scattering lengths of $J/\psi$-hadrons and 
the elastic cross sections at the threshold in the physical unit.
 For the $J/\psi$-$\pi$ channel, the data for all three volumes of
 $L=24, 32$ and 48 are used, while for the $J/\psi$-$\rho$ and 
 $J/\psi$-$N$ channels, only the data for $L=32$ and 48 are used.
 In the table, ``PSF" and ``LLE" stand for the 
 Phase Shift Formula and Leading large-$L$ Expansion, respectively.
 ``SAV" stands for the spin-averaged value,
 $\frac{1}{9}[5(a_0)_{2}+3(a_0)_{1}+(a_0)_{0}]$,
 for the $J/\psi$-$\rho$ channel
 and $\frac{1}{3}[2(a_0)_{3/2}+(a_0)_{1/2}]$ for the $J/\psi$-$N$ channel.
}
\begin{ruledtabular}
\begin{tabular}{cc|cc|cc}
& & From PSF     &  & From LLE  &     \\ 
Channel  & Spin       & $a_{0}$ [fm]  & $\sigma_{\rm el}$ [mb]  & $a_{0}$ [fm]  & $\sigma_{\rm el}$ [mb] \\ \hline\hline
$J/\psi$-$\pi$   & 1   & 0.0119$\pm$0.0039 & ${0.018}^{+0.013}_{-0.010}$ 
                                    & 0.0119$\pm$0.0025 & ${0.018}^{+0.008}_{-0.007}$  \\ \hline
$J/\psi$-$\rho$& 0   & 0.32$\pm$0.12 & ${12.9}^{+11.0}_{-7.6}$  
                                    & 0.23$\pm$0.06 & ${6.6}^{+4.2}_{-3.2}$  \\
                 & 1             & 0.25$\pm$0.10 & ${7.9}^{+7.3}_{-4.9}$   
                                    & 0.19$\pm$0.06 & ${4.6}^{+3.3}_{-2.4}$   \\
                 & 2             & 0.21$\pm$0.09 & ${5.5}^{+5.6}_{-3.7}$    
                                    & 0.17$\pm$0.06 & ${3.5}^{+2.8}_{-2.0}$   \\
                 & SAV
                 		    & 0.23$\pm$0.08 & ${6.8}^{+6.1}_{-4.2}$                   
                                    & 0.18$\pm$0.05 & ${4.1}^{+2.9}_{-2.1}$
                                    \\\hline
$J/\psi$-$N$ & 1/2   & 0.57$\pm$0.42 & ${41}^{+83}_{-38}$      
                                    & 0.35$\pm$0.15 & ${15}^{+15}_{-10}$ \\ 
                 & 3/2         & 0.88$\pm$0.63 & ${96}^{+188}_{-89}$     
                                   & 0.43$\pm$0.16& ${23}^{+20}_{-14}$ \\
                 & SAV & 0.71$\pm$0.48 & ${64}^{+116}_{-57}$
                                      & 0.39$\pm$0.14   & ${20}^{+16}_{-12}$ \\                
\end{tabular}
\end{ruledtabular}
\label{Table_SLphys}
\end{table}
%

%
%
\begin{table}[htdp]
\caption{Fitting parameter $A$ as  
the coefficient of the leading term in the large-$L$ expansion of the energy shift.
For the $J/\psi$-$\pi$ channel, we used all energy shifts at $L=24$, 32 and 48 
to extract the fitting parameter $A$. 
On the other hand, for the $J/\psi$-$\rho$ and $J/\psi$-$N$ channels, 
we excluded the energy shifts at $L=24$ for evaluating the parameters $A$.
}
\begin{ruledtabular}
\begin{tabular}{ccccccc}
&
$ A_{J/\psi \textit{\rm -}\pi}$ &
$ A_{J/\psi \textit{\rm -}\rho}^{0}$ &
$ A_{J/\psi \textit{\rm -}\rho}^{1}$ &
$ A_{J/\psi \textit{\rm -}\rho}^{2}$ &
$ A_{J/\psi \textit{\rm -}N}^{1/2}$ &
$ A_{J/\psi \textit{\rm -}N}^{3/2}$ 
 \\\hline
phys.&-24(5) &-101(28)&-84(26)&-73(25) &-115(48)&-140(52)\\
\end{tabular}
\end{ruledtabular}
\label{Table_Ampl}
\end{table}%


%
%
\begin{table}[htdp]
\caption{Results of the chiral extrapolated values of 
$s$-wave scattering lengths in lattice units for all $\eta_c$-hadron channels.
 The labels are the same as in Table~\ref{Table:EsftL24}.
}
\begin{ruledtabular}
\begin{tabular}{cc|c|c|c}
$L^3\times T$ 
&&
$ a_0^{\eta_c \textit{\rm -}\pi}$ &
$ a_0^{\eta_c \textit{\rm -}\rho}$ &
$ a_0^{\eta_c \textit{\rm -}N}$ 
 \\\hline
$24^3\times 48$ &
phys.& 0.15(5)             &-0.25(48)&1.10(1.29)\\
&chiral& -0.0025(15)&-0.25(48)&1.09(1.28) \\\hline
$32^3\times 48$ &
phys.& 0.15(5)          &2.8(1.6)&8.8(9.2)\\
&chiral& -0.0013(9)&2.7(1.6)&8.7(9.0) \\\hline
$48^3\times 48$ &
phys.& 0.33(17)             &3.4(1.8)&12.4(10.5)\\
&chiral& -0.0014(14)&3.4(1.8)&12.2(10.2) \\\hline
\end{tabular}
\end{ruledtabular}
\label{Table_SL_etac}
\end{table}
%

%
%
\begin{table}[htdp]
\caption{The $s$-wave scattering lengths of $\eta_c$-hadrons and 
the elastic cross sections at the threshold in the physical unit.
For the $\eta_c$-$\pi$ channel, the data for all three volumes of
$L=24, 32$ and 48 are used, while for the $\eta_c$-$\rho$ and 
$\eta_c$-$N$ channels, only the data for $L=32$ and 48 are used.
The labels are the same as in Table~\ref{Table_SLphys}.
}
\begin{ruledtabular}
\begin{tabular}{cc|cc|cc}
& & From PSF     &  & From LLE  &     \\ 
Channel  & Spin       & $a_{0}$ [fm]  & $\sigma_{\rm el}$ [mb]  & $a_{0}$ [fm]  & $\sigma_{\rm el}$ [mb] \\ \hline\hline
$\eta_c$-$\pi$   & 0   & 0.0113$\pm$0.0035 & ${0.016}^{+0.011}_{-0.008}$  
                                    & 0.0112$\pm$0.0024 & ${0.016}^{+0.008}_{-0.006}$  \\ \hline
$\eta_c$-$\rho$& 1   & 0.21$\pm$0.11 & ${5.3}^{+7.5}_{-4.3}$ 
                                    & 0.16$\pm$0.05 & ${3.4}^{+2.5}_{-1.8}$    \\\hline
$\eta_c$-$N$ & 1/2   & 0.70$\pm$0.66 & ${62}^{+172}_{-62}$     
                                   & 0.39$\pm$0.14& ${19}^{+16}_{-11}$  \\
\end{tabular}
\end{ruledtabular}
\label{Table_SLphys_etac}
\end{table}
%

%
%
\begin{table}[htdp]
\caption{Fitting parameter $A$ as 
the coefficient of the leading term in the large-$L$ expansion of the energy shift.
For the $\eta_c$-$\pi$ channel, we use all energy shifts at $L=24$, 32 and 48 to extract 
the fitting parameter $A$. On the other hand, for the $\eta_c$-$\rho$ and $\eta_c$-$N$ channels, 
we exclude the energy shifts at $L=24$ for evaluating the parameters $A$.
}
\begin{ruledtabular}
\begin{tabular}{cccc}
 & 
$ A_{\eta_c \textit{\rm -}\pi}$ &
$ A_{\eta_c \textit{\rm -}\rho}$ &
$ A_{\eta_c \textit{\rm -}N}$ 
 \\\hline
phys.&-23(5) &-72(23)&-127(47)\\
\end{tabular}
\end{ruledtabular}
\label{Table_Ampl_etac}
\end{table}%

%
%
%

%
%
%

%
%
%

%
%
%

%
%
%

%
%

\begin{thebibliography}{99}
\bibitem{footnote1}  
In Ref.~\cite{YSHH}, 
we had only results on $L=24$ and $L=32$
lattices. Therefore, we could not rule out possibility of bound state formation in
$J/\psi$-$N$ channels since the volume dependence from $L=24$ to $L=32$
has an opposite tendency against that of scattering states. The $L=48$ data
shown in the present paper could rule out this possibility.
\end{thebibliography}
\end{document}